\documentclass[man,12pt,nolmodern,floatsintext,natbib]{apa7}
\usepackage[nodisplayskipstretch]{setspace}
\usepackage{booktabs,multirow}
\usepackage{graphicx}
\usepackage{enumitem}
\usepackage{natbib}
\usepackage{amsmath,amsfonts,amsthm}
\usepackage{algorithm,algpseudocode}
\usepackage{tikz}
\usetikzlibrary{shapes,arrows.meta,decorations.markings,positioning}

\DeclareRobustCommand{\bfseries}{\fontseries{b}\selectfont}
\DeclareMathAlphabet{\mathbf}{OT1}{cmr}{b}{n}

\definecolor{red1}{RGB}{228,55,55}
\definecolor{blue1}{RGB}{0,0,0}
\long\def\yl#1{\bgroup\color{blue1}#1\egroup}

\def\t{^\top}

\def\aalpha{\boldsymbol{\alpha}}
\def\bbeta{\boldsymbol{\beta}}

\def\ttheta{\boldsymbol{\theta}}

\def\pphi{\boldsymbol{\varphi}}
\def\ppsi{\boldsymbol{\psi}}

\def\kkappa{\boldsymbol{\kappa}}
\def\llambda{\boldsymbol{\lambda}}

\def\XXi{\mathbf{\Xi}}

\def\BB{\mathbf{B}}
\def\DD{\mathbf{D}}
\def\EE{\mathbf{E}}
\def\II{\mathbf{I}}

\def\YY{\mathbf{Y}}

\def\yy{\mathbf{y}}

\def\real{\mathbb{R}}

\def\vec{\hbox{vec}}

\def\var{\hbox{Var}}

\def\ex{\mathbb{E}}
\def\pr{\mathrm{P}}

\def\cX{\mathcal{X}}
\def\cY{\mathcal{Y}}

\newtheoremstyle{default}
  {0pt} % Space above
  {0pt} % Space below
  {\itshape} % Body font
  {} % Indent amount
  {\bfseries} % Theorem head font
  {.} % Punctuation after theorem head
  {.5em} % Space after theorem head
  {} % Theorem head spec (can be left empty, meaning `normal')
\newtheoremstyle{remark}
  {0pt} % Space above
  {0pt} % Space below
  {} % Body font
  {} % Indent amount
  {\itshape} % Theorem head font
  {.} % Punctuation after theorem head
  {.5em} % Space after theorem head
  {} % Theorem head spec (can be left empty, meaning `normal')
\theoremstyle{default}

\theoremstyle{remark}

\algrenewcommand\algorithmicrequire{\textbf{input:}}
\algrenewcommand\algorithmicensure{\textbf{output:}}
\algblockdefx{Do}{EndDo}{\textbf{do}}{\textbf{end do}}

\setlist[itemize, 1]{leftmargin=*, topsep=0ex, itemsep=0pt, parsep=0ex, labelindent=0pt}
\setlist[itemize, 2]{leftmargin=*, topsep=0ex, itemsep=0pt, parsep=0ex, labelindent=0pt, label=$\circ$}
\setlist[itemize, 3]{leftmargin=*, topsep=0ex, itemsep=0pt, parsep=0ex, labelindent=0pt, label=-}
\setlist[enumerate, 1]{leftmargin=*, topsep=0ex, itemsep=0pt, parsep=0ex, labelindent=0pt, label=(\arabic*)}

\begin{document}

% title page
\title{What Can We Learn from a Semiparametric Factor Analysis of Item Responses and Response Time? An Illustration with the PISA 2015 Data}

\shorttitle{PISA Responses and Response Time}
\authorsnames[1,1]{Yang Liu, Weimeng Wang}
\authorsaffiliations{{Department of Human Development and Quantitative Methodology\\ University of Maryland, College Park}}
\authornote{Correspondence should be made to Yang Liu at
3304R Benjamin Bldg, 3942 Campus Dr, University of Maryland, College Park, MD
20742. Email: yliu87@umd.edu. The dataset analyzed during the current study is available in the OECD PISA Database (https://www.oecd.org/pisa/data/). The work is sponsored by the National Science Foundation under grant No. 1826535. The authors declare that they have no known competing financial interests or personal relationships that could have appeared to influence the work reported in this paper.}

% abstract and keywords
\abstract{%
\indent
It is widely believed that a joint factor analysis of item responses and response time (RT) may yield more precise ability scores that are conventionally predicted from responses only. For this purpose, a simple-structure factor model is often preferred as it only requires specifying an additional measurement model for item-level RT while leaving the original item response theory (IRT) model for responses intact. The added speed factor indicated by item-level RT correlates with the ability factor in the IRT model, allowing RT data to carry additional information about respondents' ability. However, parametric simple-structure factor models are often restrictive and fit poorly to empirical data, which prompts under-confidence in the suitablity of a simple factor structure. In the present paper, we analyze the 2015 Programme for International Student Assessment (PISA) mathematics data using a semiparametric simple-structure model. We conclude that a simple factor structure attains a decent fit after further parametric assumptions in the measurement model are sufficiently relaxed. Furthermore, our semiparametric model implies that the association between latent ability and speed/slowness is strong in the population, but the form of association is nonlinear. It follows that scoring based on the fitted model can substantially improve the precision of ability scores.
}
\maketitle
\setcounter{secnumdepth}{2}

\section{Introduction}
\label{s:intro}
Psychometric investigation on cognitive ability and speed has a long and rich history \citep[e.g.,][]{Carroll1993, Gulliksen1950, Luce1986, ThorndikeEtAl1926}. In the \citeyear{ThorndikeEtAl1926} monograph, \citeauthor{ThorndikeEtAl1926} stated that ``level'', ``extent'', and ``speed'' are three distinct aspects in any measure of performance: While both ``level'' and ``extent'' are manifested by correctness of answers and thus can be collectively translated to ability in modern terminology, ``the speed of producing any given product is defined, of course, by the time required'' \citep[][p. 26]{ThorndikeEtAl1926}. The prevalence of computerized test administration and data collection in recent years facilitates the acquisition of response-time (RT) data at the level of individual test items. In parallel, we witnessed a mushrooming development of psychometric models for item responses and RT over the past few decades \citep[see][for reviews]{DeBoeckJeon2019, Goldhammer2015}, which in turn gave rise to broader investigations on the relationship between response speed and accuracy in various substantive domains \citep[see][for reviews]{LeeChen2011, KyllonenZu2016, vonDavierEtAl2019}. Empirical findings suggested that response speed not only composes proficiency or informs the construct to be measured but also bespeaks secondary test-taking behaviors such as rapid guessing \citep{DeriboEtAl2021, Wise2017}, using preknowledge \citep{QianEtAl2016, Sinharay2020, SinharayJohnson2020}, lacking motivation \citep{Finn2015, Thurstone1937, WiseKong2005}, etc.

Characterizing individual differences in ability and speed with item responses and RT data is in essence a factor analysis problem \citep{MolenaarEtAl2015a, MolenaarEtAl2015b}. The two-factor simple-structure model proposed by \citet{vanderLinden2007} was arguably the most popular modeling option so far: Item responses and log-transformed RT variables are treated as two independent clusters of observed indicators for the ability and speed/slowness factors, respectively, and the two latent factors jointly follow a bivariate normal distribution (see Figure 2 of \citet{MolenaarEtAl2015b} for a path-diagram representation). A notable merit of the simple-structure factor model is its plug-and-play nature: \yl{Analysts can separately apply standard item response theory (IRT) models for discrete responses \cite[e.g., one-, two-, three-, or four-parameter logistic (1-4PL) model;][]{Birnbaum1968, BartonLord1981} and standard factor analysis models for the continuous log-RT variables \cite[e.g., linear-normal factor model;][]{Joreskog1969}, and then simply let the two latent factors covary}. Despite its succinctness and popularity, the simple-structure model may fit poorly to empirical data. A highly endorsed interpretation for the lack of fit is that the two inter-dependent latent factors cannot fully explain the dependencies among item-level responses and RT variables. Based on this rationale, numerous diagnostics for residual dependencies and remedial modifications of the simple-structure model have been proposed in the recent literature \citep[e.g.,][]{BolsinovaEtAl2017a, BolsinovaMaris2016, BolsinovaMolenaar2018, BolsinovaEtAl2017b, BolsinovaTijmstra2016, GlasvanderLinden2010, MengEtAl2015, RangerOrtner2012, vanderLindenGlas2010}. 

Augmenting standard IRT models with a measurement component for item-level RT may result in more precise ability scores, which is often highlighted as a practical benefit of RT modeling in educational assessment \citep{BolsinovaTijmstra2018, vanderLindenEtAl2010}. Under a simple-structure model with bivariate normal factors, the degree to which item-level RT improves scoring precision is dictated by the strength of the inter-factor correlation \citep[see Study 1 of][]{vanderLindenEtAl2010}. However, near-zero correlation estimates between ability and speed were sometimes encountered in real-world applications \citep[e.g.,][]{BolsinovaEtAl2017a, BolsinovaEtAl2017b, LeeJia2014, vanderLindenEtAl1999}. Whenever it happens, analysts are inclined to conclude that item-level RT is not useful for ability estimation at all, or that a less parsimonious factor structure is needed to enhance the utility of RT for scoring purposes \citep[e.g., allowing the log-RT variables to cross-load on the ability factor][]{BolsinovaTijmstra2018}.

Indeed, \citeauthor{vanderLinden2007}'s \citeyearpar{vanderLinden2007} model could be overly restrictive for analyzing item responses and RT data. We, however, do not want to rush to the conclusion that it is the simple factor structure that should be blamed and abandoned. \yl{Other parametric assumptions, such as link functions, linear or curvilinear dependencies, and distributions of latent traits and error terms, are also part of the model specification and may contribute to the misfit as well.} A fair evaluation on the tenability and usefulness of a simple factor structure demands a version of the model with minimal parametric assumptions other than the simple factor structure itself, which we refer to as a {\it semiparametric simple-structure model}. Should the semiparametric model still struggle to fit the data adequately, we no longer hesitate to give up on the simple factor structure.

Fortunately, the major components of a semiparametric simple-structure factor analysis have been readily developed in the existing literature. They are
\begin{enumerate}
  \item a semiparametric (unidimensional) IRT model for dichotomous and polytomous responses \citep{AbrahamowiczRamsay1992, RossiEtAl2002}; 
  \item a semiparametric (unidimensional) factor model for continuous log-RT variables \citep{LiuWang2022}
  \item a nonparametric copula density estimator for ability and speed/slowness with fixed marginals \citep{KauermannEtAl2013, DouEtAl2021}.
\end{enumerate}
\yl{As a side remark, we are aware of alternative semiparametric approaches that can be used for each of the above three components: for example, the monotonic polynomial logistic model for item responses \citep{FalkCai2016a, FalkCai2016b}, the proportional hazard model \citep{Kang2017, RangerKuhn2012, WangEtAl2013b} and the linear transformation model \citep{WangEtAl2013a} for item-level RT, and the finite normal mixture model \citep{Bauer2005, PekEtAl2009} and the Davidian curve model \citep{WoodsLin2009, ZhangDavidian2001, ZhangEtAl2021} for the joint distribution of latent traits. However, we focus on methods based on smoothing splines in the current analysis. Besides, the simultaneous incorporation of flexible models for all the three components of a simple structure model appears to be novel in the literature of RT modeling. Compared to, e.g., \citet{WangEtAl2013a} and \citet{WangEtAl2013b}, in which semiparametric models were applied to only the RT data, our model fares more flexible and thus is more likely to reveal sophisticated dependency patterns in a joint analysis of item responses and RT data.
} 

By retrospectively analyzing a set of mathematics testing data from the 2015 Programme for International Student Assessment \citep[PISA;][]{OECD2016}, we revisit the following research questions that have only been partially answered previously through parametric simple-structure models:
\begin{enumerate}
  \item Is a simple factor structure sufficient for a joint analysis of item response and RT?
  \item How strong are math ability and general processing speed associated in the population of respondents?
  \item To what extent can processing speed improve the precision in ability estimates under a simple-structure model?
\end{enumerate}
It is worth mentioning that the data set was previously analyzed by \citet{ZhanEtAl2018} using a variant of \citeauthor{vanderLinden2007}'s \citeyearpar{vanderLinden2007} simple-structure model with testlet effects: A higher-order cognitive diagnostics model with testlet effects was used for item responses, a linear-normal factor model was used for log-transformed RT, and the (higher-order) ability and speed factors were assumed to be bivariate normal. \citet{ZhanEtAl2018} reported an estimated inter-factor correlation of $-0.2$ and hence concluded that the association between speed and ability is weak. We are particularly interested in whether their conclusion stands after abandoning inessential parametric assumptions other than the simple factor structure.

The rest of the paper is organized as follows. We first provide a technical introduction of the proposed semiparametric procedure in Section \ref{s:met}: The three components of the semiparametric simple-structure model are formulated in Sections \ref{ss:1dim} and \ref{ss:cop}, penalized maximum likelihood (PML) estimation and empirical selection of penalty weights are outlined in Section \ref{ss:est}, and bootstrap-based goodness-of-fit assessment and inferences are described in Sections \ref{ss:fit} and \ref{ss:lv}. Descriptive statistics for the 2015 PISA mathematics data and a plan of our analysis are summarized in Section \ref{s:dap}, followed by a detailed report of results in Section \ref{s:res}. The paper concludes with a discussion of broader implications of our findings and limitations of our method.

\section{Methods}
\label{s:met}

\subsection{Unidimensional Semiparametric Factor Models}
\label{ss:1dim}

Let $Y_{ij}\in\cY_j\subset\real$ be the $j$th \textit{manifest variable (MV)} observed for respondent $i$: $Y_{ij}$ represents either a discrete response to a test item or a continuous item-level RT. In our semiparametric factor model, the distribution of $Y_{ij}$ is characterized by the following logistic conditional density\footnote{For simplicity, both probability density functions for continuous random variables and probability mass functions for discrete random variables are referred to as densities.} of $Y_{ij} = y\in\cY_j$ given a unidimensional \textit{latent variable} (LV; also known as latent factor, latent trait, etc.) $X_i = x\in\cX\subset\real$:
\begin{equation}
  f_j(y|x) = \frac{\exp\left(g_j(x, y)\right)}{\int_{\cY_j}\exp\left(g_j(x, y')\right)\mu_j(dy')},
  \label{eq:cdns}
\end{equation}
in which the normalizing integral with respect to the dominating measure $\mu_j$ on $\cY_j$ is assumed to be finite. Equation \ref{eq:cdns} defines a valid conditional density as it is non-negative and integrates to unity with respect to $y$ for a given $x$. However, the bivariate function $g_j: \cX\times\cY_j\to\real$ is not identifiable: It is not difficult to see that adding any univariate function of $x$ to $g_j(x, y)$ does not change the value of Equation \ref{eq:cdns} \citep{Gu1995, Gu2013}. To impose necessary identification constraints, we re-write $g_j$ by the functional analysis of variance (fANOVA) decomposition
\begin{equation}
  g_j(x, y) = g_j^y(y) + g_j^{xy}(x, y)
  \label{eq:fanova}
\end{equation}
and require that 
\begin{equation}
  g_j^y(y_0) = 0,\ g_j^{xy}(x_0, y)\equiv 0,\ \hbox{and}\ g_j^{xy}(x, y_0)\equiv 0
  \label{eq:side}
\end{equation}
for some reference levels $x_0\in\cX$ and $y_0\in\cY_j$. Equation \ref{eq:side} is referred to as \textit{side conditions}; $x_0$ and $y_0$ can be set arbitrarily within the respective domains \citep[see][for more detailed comments]{LiuWang2022}. The univariate component $g_j^y$ and the bivariate component $g_j^{xy}$ are functional parameters to be estimated from observed data.

Let $\ppsi_j: \cY_j\to\real^{L_j}$ be a collection of $L_j$ basis functions defined on the support of $Y_{ij}$, and $\pphi: \cX\to\real^K$ be a collection of $K$ basis functions defined on the support of $X_i$. We proceed to approximate the functional parameters by basis expansion. In particular, we set the univariate component
\begin{equation}
  g_j^y(y) = \ppsi_j(y)\t\aalpha_j,
  \label{eq:approx1}
\end{equation}
in which the coefficient vector $\aalpha_j\in\real^{L_j}$ satisfies 
\begin{equation}
  \ppsi_j(y_0)\t\aalpha_j = 0.
  \label{eq:constr1}
\end{equation}
Similarly, the bivariate component is expressed as 
\begin{equation}
  g_j^{xy}(x, y) = \ppsi_j(y)\t\BB_j\pphi(x),
  \label{eq:approx2}
\end{equation}
in which the coefficient matrix $\BB_j\in\real^{L_j\times K}$ satisfies 
\begin{equation}
  \BB_j\pphi(x_0) = {\bf 0}\ \hbox{and}\ \BB_j\t\ppsi_j(y_0) = {\bf 0}. 
  \label{eq:constr2}
\end{equation}
The linear constraints imposed for the coefficients $\aalpha_j$ and $\BB_j$ (Equations \ref{eq:constr1} and \ref{eq:constr2}) guarantee that the side conditions (Equation \ref{eq:side}) are satisfied. 

\paragraph{Continuous Data}
When both $X_i\in\cX$ and $Y_{ij}\in\cY_j$ (equipped with the Lebesgue measure $\mu_j$) are continuous random variables defined on closed intervals, Equation \ref{eq:cdns} corresponds to the semiparametric factor model considered by \citet{LiuWang2022}. Without loss of generality, let $\cX = \cY_j = [0, 1]$. In fact, any closed interval can be rescaled to the unit interval via a linear transform: If $z\in[a, b]$, $a < b$, then $(z - a)/(b - a)\in[0, 1]$. To approximate smooth functional parameters supported on unit intervals or squares, we use the same cubic B-spline basis with equally spaced knots \citep{DeBoor1978} for both $\ppsi_j$ and $\pphi$ (and thus $L_j = K$). It is sometimes desirable to force the MV to be stochastically increasing as the LV increases. \citet{LiuWang2022} considered a simple approach to impose likelihood-ratio monotonicity, which boils down to the following linear inequality constraints on the coefficient matrix $\BB_j$:
\begin{equation}
  (\DD_K\otimes\DD_K)\,\vec(\BB_j)\ge{\bf 0}.
  \label{eq:mono}
\end{equation}
In Equation \ref{eq:mono}, $\vec(\cdot)$ denotes the vectorization operator, and 
\begin{equation*}
  \DD_K = \begin{bmatrix}
    1 & -1 & & \\
    & \ddots & \ddots & \\
    & & 1 & -1
  \end{bmatrix}
\end{equation*}
is a $(K - 1)\times K$ first-order difference matrix. We also set $\DD_1 = 1$ by convention.

\paragraph{Discrete Data}
When $\cY_j = \{0, \dots, C_j - 1\}$ and $\mu_j$ is the associated counting measure, let $y_0 = 0$, $L_j = C_j - 1$, and $\ppsi_j(y) = (\psi_{j1}(y), \dots, \psi_{j,C_j-1}(y))\t$ such that $\psi_{jk}(y) = 1$ if $y = k$ and 0 if $y\ne k$. Then our generic model (Equations \ref{eq:cdns} and \ref{eq:fanova}) reduces to \citeauthor{AbrahamowiczRamsay1992}'s \citeyearpar{AbrahamowiczRamsay1992} multi-categorical semiparametric IRT model for unordered polytomous responses, which is further equivalent to the semiparametric logistic IRT proposed by \citet{RamsayWinsberg1991} and \citet{RossiEtAl2002} when $C_j = 2$ (i.e., dichotomous data). It is because the  basis expansions (i.e., Equations \ref{eq:approx1} and \ref{eq:approx2}) are simplified to $g_j^y(y) = \alpha_{jy}$ and $g_j^{xy}(x, y) = \pphi(x)\t\bbeta_{jy}$, in which $\bbeta_{jy}\t$ denotes the $y$th row of $\BB_j$, if $y = 1, \dots, C_j - 1$; meanwhile, $g_j^y(0) = 0$ and $g_j^{xy}(x, 0)\equiv 0$ as part of the side conditions. The conditional density (e.g., Equation \ref{eq:cdns}) then becomes the item response function (IRF)
\begin{equation}
  f_j(y|x) = \left\{\begin{array}{ll}
      \displaystyle\frac{1}{1 + \sum_{c = 1}^{C_j - 1}\exp\left(\alpha_{jc} + \pphi(x)\t\bbeta_{jc}\right)}, & \hbox{if }y = 0,\\[20pt]
      \displaystyle\frac{\exp\left(\alpha_{jy} + \pphi(x)\t\bbeta_{jy}\right)}{1 + \sum_{c = 1}^{C_j - 1}\exp\left(\alpha_{jc} + \pphi(x)\t\bbeta_{jc}\right)}, & \hbox{if }y = 1,\dots, C_j - 1.
\end{array}\right.
  \label{eq:irf}
\end{equation}
Like the continuous case, we only consider $\cX = [0, 1]$ and $\pphi$ being a cubic B-spline basis defined by a sequence of equally spaced knots. Similar to Equation \ref{eq:mono} in the continuous case, we may impose likelihood-ratio monotonicity on the conditional density by
\begin{equation}
  (\DD_K\otimes\DD_{C_j - 1})\,\vec(\BB_j)\ge{\bf 0},
  \label{eq:monod}
\end{equation}
which reduces to $\DD_k\bbeta_{j1}\ge{\bf 0}$ when $C_j = 2$ (i.e., dichotomous items).

\subsection{Simple Factor Structure and Latent Variable Density}
\label{ss:cop}
Consider a battery of \yl{$m_1$ continuous MVs and $m_2$ discrete MVs} and write $m = m_1 + m_2$. We typically have $m_1 = m_2 = m/2$ when the discrete responses and continuous RT variables are observed for the same set of items. From now on, denote by $Y_{i1}, \dots, Y_{i,m_1}$ the base-10 log-transformed RT, each of which is rescaled to $[0, 1]$, and by $Y_{i, m_1+1}, \dots, Y_{im}$ the corresponding responses. Let $X_{i1}, X_{i2}\in[0, 1]$ be the \textit{slowness}\footnote{Slowness is the reversal of speed. We abide by the convention that the LV is positively associated with the MV.} and \textit{ability} factors for respondent $i$, respectively. A \textit{simple factor structure} requires that the item responses $Y_{i,m_1+1}, \dots, Y_{im}$ are conditionally independent of the slowness factor $X_{i1}$ given the ability factor $X_{i2}$, and symmetrically that the log-RT variables $Y_{i1}, \dots, Y_{i,m_1}$ are independent of $X_{i2}$ given $X_{i1}$. We also make the \textit{local independence} assumption that is standard in factor analysis \citep{McDonald1982}: $Y_{i1}, \dots, Y_{im}$ are mutually independent conditional on $X_{i1}$ and $X_{i2}$. Further let $\YY_i = (Y_{i1}, \dots, Y_{im})\t$ collect all the MVs produced by respondent $i$. The simple structure and local independence assumptions imply that
\begin{equation}
  f(\yy|x_1, x_2) = \prod_{j=1}^{m_1}f(y_j|x_1)\cdot\prod_{j=m_1+1}^{m}f(y_j|x_2),
  \label{eq:jointcond}
\end{equation}
in which $\yy = (y_1, \dots, y_m)\t\in[0, 1]^{m_1}\times\cY_{m_1 + 1}\times\cdots\times\cY_{m}$, and $x_1, x_2\in[0, 1]$.

For convenience in approximating functional parameters, both $X_{i1}$ and $X_{i2}$ are assumed to follow a Uniform[0, 1] distribution marginally. However, we are aware that uniformly distributed LVs are less attractive for substantive interpretation. Adopting the strategy of \citet{LiuWang2022}, we define $X_{id}^* = \Phi^{-1}(X_{id})$, $d = 1, 2$, where $\Phi^{-1}$ is the standard normal quantile function; the transformed LVs are marginally ${\cal N}(0, 1)$ variates, in agreement with the standard formulation in parametric factor analysis. To capture the potentially complex association between latent slowness and ability, we employ a nonparametric estimator for the copula density \citep{Sklar1959, Nelsen2006} of $(X_{i1}, X_{i2})\t$, denoted $c(x_1, x_2)$. A copula density is non-negative and has uniform marginals: That is,
\begin{equation}
  c(x_1, x_2)\ge 0\ \hbox{and }\int_0^1 c(x_1, x_2)dx_1 = \int_0^1 c(x_1, x_2)dx_2\equiv 1,\ \forall x_1, x_2\in[0, 1].
  \label{eq:copula}
\end{equation}
$c$ is in fact the joint density of $(X_{i1}, X_{i2})\t$ since both $X_{i1}$ and $X_{i2}$ are marginally uniform. In the light of Sklar's theorem, the joint density of the transformed $(X_{i1}^*, X_{i2}^*)\t$ can be calculated by
\begin{equation}
  h(x_1^*, x_2^*) = c(\Phi(x_1^*), \Phi(x_2^*))\phi(x_1^*)\phi(x_2^*),
  \label{eq:jdnstrans}
\end{equation}
in which $\phi$ and $\Phi$ are the density and distribution functions of ${\cal N}(0, 1)$, respectively.

We approximate the bivariate copula density $c$ by a tensor-product spline \citep{DouEtAl2021, KauermannEtAl2013}:
\begin{equation}
  c(x_1, x_2) = \pphi(x_2)\t\XXi\pphi(x_1)
  \label{eq:bsplcop}
\end{equation}
in which $\pphi: [0, 1]\to\real^K$ is a set of cubic B-spline basis functions defined with equally spaced knots\footnote{For simplicity, the same set of basis functions is used for the LVs in Equations \ref{eq:approx2}, \ref{eq:irf}, and \ref{eq:bsplcop}.}, and $\XXi$ is an $K\times K$ coefficient matrix. For Equation \ref{eq:bsplcop} to be a proper copula density, we impose the following linear constraints on $\XXi$:
\begin{equation}
  \xi_{kl}\ge 0,\ \forall k, l = 1, \dots, K,\ \hbox{and }\XXi\kkappa = \XXi\t\kkappa = {\bf 1}
  \label{eq:constrcop},
\end{equation}
in which $\xi_{kl}$ is the $(k, l)$th element of $\XXi$, and 
\begin{equation}
  \kkappa = \int_0^1\pphi(x)dx
  \label{eq:nccop}
\end{equation}
is a $K\times 1$ vector of normalizing constants for basis functions. It can be verified by elementary properties of B-splines and straightforward algebra that Equations \ref{eq:bsplcop} and \ref{eq:constrcop} imply Equation \ref{eq:copula}.

\subsection{Estimation}
\label{ss:est}
For each MV $j = 1,\dots, m$, let $\ttheta_j = (\aalpha_j\t, \vec(\BB_j)\t)\t$ collect all the coefficients in $g_j^y$ and $g_j^{xy}$. Also let $\ttheta = (\ttheta_j\t, \dots, \ttheta_m\t, \vec(\XXi)\t)\t$ denote all the coefficients in the simple-structure factor model. We estimate $\ttheta$ by \textit{penalized maximum (marginal) likelihood (PML)}.  The marginal likelihood for the MV vector $\YY_i = \yy$ amounts to the integration of Equation \ref{eq:jointcond} over $x_1$ and $x_2$ under the copula density $c(x_1, x_2)$: That is,
\begin{equation}
  f(\yy; \ttheta) = \iint_{[0, 1]^2}f(\yy|x_1, x_2)c(x_1, x_2)dx_1dx_2.
  \label{eq:mlik}
\end{equation}
Pooling across an independent and identically distributed (i.i.d.) sample of size $n$, we arrive at the sample log-likelihood function
\begin{equation}
  \ell(\ttheta; \yy_{1:n}) = \sum_{i=1}^n\log f(\yy_i; \ttheta),
  \label{eq:loglikn}
\end{equation}
in which $\yy_{1:n} = (\yy_1, \dots, \yy_n)\t$ denotes an $n\times m$ matrix of observed MV data.

To avoid overfitting, we regularize the roughness of estimated functional parameters by quadratic-form penalties in spline coefficients. For a continuous MV $j$, the penalty term for $\ttheta_j$ is the sum of a univariate P-spline penalty for $\aalpha_j$ and a bivariate P-spline penalty for $\BB_j$ \citep{EilersMarx1996, CurrieEtAl2006}:
\begin{equation}
  q_j(\ttheta_j; \lambda_j) = \frac{\lambda_j}{2}\aalpha_j\t\EE_K\t\EE_K^{}\aalpha_j^{}
  + \frac{\lambda_j}{2}\vec(\BB_j)\t\left( \II_K\otimes\EE_K\t\EE_K^{} + \EE_K\t\EE_K^{}\otimes\II_K \right)\vec(\BB_j),
  \label{eq:penitemc}
\end{equation}
in which $\lambda_j > 0$ is the \textit{penalty weight}, $\II_K$ denotes a $K\times K$ identity matrix, and 
\begin{equation*}
  \EE_K = \begin{bmatrix}
    1 & -2 & 1 & & \\
      & \ddots & \ddots & \ddots &\\
    && 1 & -2 & 1 \\
  \end{bmatrix}
\end{equation*}
is a second-order difference matrix of dimension $(K - 2)\times K$. If the MV is polytomous, no penalty is needed for the intercepts $\aalpha_j$ and columns of $\BB_j$. The resulting P-spline penalty term then becomes
\begin{equation}
  q_j(\ttheta_j; \lambda_j) = \frac{\lambda_j}{2}\vec(\BB_j)\t\left(\EE_K\t\EE_K^{}\otimes\II_{C_j - 1} \right)\vec(\BB_j).
  \label{eq:penitemd}
\end{equation}
A similar bivariate P-spline penalty is also introduced for the coefficient matrix $\XXi$:
\begin{equation}
  q(\XXi; \lambda_{m + 1}) = \frac{\lambda_{m+1}}{2}\vec(\XXi)\t\left( \II_K\otimes\EE_K\t\EE_K^{} + \EE_K\t\EE_K^{}\otimes\II_K \right)\vec(\XXi)
  \label{eq:pengrp}
\end{equation}
with a positive penalty weight $\lambda_{m+1}$. Combining Equations \ref{eq:loglikn}--\ref{eq:pengrp}, we express the penalized sample log-likelihood function as
\begin{equation}
  p(\ttheta; \yy_{1:n}, \llambda) = \ell(\ttheta; \yy_{1:n}) - n\left[\sum_{j=1}^mq_j(\ttheta_j, \lambda_j) + q(\XXi; \lambda_{m+1})\right],
  \label{eq:ploglikn}
\end{equation}
in which $\llambda = (\lambda_1, \dots, \lambda_m, \lambda_{m+1})\t\in(0, \infty)^{m + 1}$. PML estimation amounts to finding $\ttheta$ that maximizes Equation \ref{eq:ploglikn} subject to a series of linear equality and inequality constraints (i.e., Equations \ref{eq:constr1}, \ref{eq:constr2}, \ref{eq:mono}, \ref{eq:monod}, and \ref{eq:constrcop}), which is accomplished by a modified expectation-maximization \cite[EM;][]{BockAitkin1981, DempsterEtAl1977} algorithm. A sequential quadratic programming algorithm \cite[][Algorithm 18.3]{NocedalWright2006} is employed in the M-step to handle constrained optimization. The algorithm is a simple extension to what was described in Sections 4.1 and 4.2 of \cite{LiuWang2022}; further details are therefore omitted for succinctness. Denote by $\hat\ttheta(\yy_{1:n}, \llambda)$ the PML estimates of $\ttheta$ obtained from data $\yy_{1:n}$ and penalty weights $\llambda$.

Larger penalty weights enforce less variable yet more biased solutions and \textit{vice versa}---a well-known phenomenon referred to as the \textit{bias-variance trade-off}. To strike a balance, we select the optimal $\llambda$ from a pre-specified grid by multi-fold cross-validation. Let $\Omega_1, \dots, \Omega_S$ be a partition of the sample: $\bigcup_{s=1}^S\Omega_s = \{1, \dots, n\}$ and $\Omega_s\cap\Omega_{s'} = \emptyset$ for all $s\ne s'$. For each $s$, let $\Omega_s^c$ be the \textit{calibration set} and $\Omega_s$ be the \textit{validation set}, in which the superscript $c$ denotes the complement of a set. Predictive adequacy associated with a particular $\llambda$ is gauged by the \textit{empirical risk}
\begin{equation}
  R(\yy_{1:n}, \llambda) = -\frac{1}{S}\sum_{s = 1}^S\frac{1}{|\Omega_s|}\ell(\hat\ttheta(\yy_{\Omega_s^c}, \llambda); \yy_{\Omega_s}),
  \label{eq:risk}
\end{equation}
in which $|\Omega_s|$ denotes the size of $\Omega_s$, $\ell(\hat\ttheta(\yy_{\Omega_s^c}, \llambda); \yy_{\Omega_s})$ denotes the log-likelihood of the validation sub-sample evaluated at the estimated coefficients \yl{from} the calibration set. Instead of choosing $\llambda$ that minimizes Equation \ref{eq:risk} (i.e., the best solution), we adopt the ``one standard error (SE)'' heuristic \citep{ChenYang2021, HastieEtAl2009} to take into account sampling variability: We select the smoothest solution within one SE from the $\llambda$ that minimizes the empirical risk, where the SE at a specific $\llambda$ is estimated by
\begin{equation}
  \mathrm{SE}(\yy_{1:n}, \llambda) = \sqrt{\frac{1}{S-1}\sum_{s = 1}^S\left[-\frac{1}{|\Omega_s|}\ell(\hat\ttheta(\yy_{\Omega_s^c}, \llambda); \yy_{\Omega_s}) - R(\yy_{1:n}, \llambda)\right]^2}.
  \label{eq:serisk}
\end{equation}

The value $\llambda$ contains $m + 1$ elements. To alleviate the computational burden for penalty weights selection, we set $\lambda_1 = \cdots = \lambda_{m_1} = \lambda_{(c)}$ for continuous MVs, $\lambda_{m_1 + 1} = \cdots = \lambda_m = \lambda_{(d)}$ for discrete MVs, and $\lambda_{m + 1} = \lambda_{(g)}$ for the copula density of the two LVs. We also resort to a multistage workaround to select the remaining three penalty weights: (1) A unidimensional model is fitted to only the continuous MVs to find the optimal $\lambda_{(c)}$, (2) a unidimensional model is fitted to only the discrete MVs to find the optimal $\lambda_{(d)}$, and (3) a two-dimensional simple-structure model is fitted to all the MVs to find the optimal $\lambda_{(g)}$ while fixing $\lambda_{(c)}$ and $\lambda_{(d)}$ at their optimal values determined in earlier stages. The optimal weights thereby selected are denoted $\hat\llambda(\yy_{1:n})$. We then refit the model using the optimal weight and the full set of data to obtain the final solution of spline coefficients $\hat\ttheta(\yy_{1:n}, \hat\llambda(\yy_{1:n}))$.

\subsection{Model Fit Diagnostics and Inferences}
\label{ss:fit}
We quantify the sampling variability of sample statistics, including goodness of fit diagnostics and approximations to functional parameters, by bootstrapping \citep{EfronTibshirani1994, HastieEtAl2009}. Let $\bar\YY_i$ be a random sample from the collection of observed MV vectors $\{\yy_1, \dots, \yy_n\}$ such that each element is selected with probability $1/n$. Sample with replacement $n$ times and denote the resulting bootstrap sample $\bar\YY_{1:n}$. We approximate the sampling distribution of any test statistic $T(\YY_{1:n})$ by the bootstrap sampling distribution of $T(\bar\YY_{1:n})$ conditional on $\yy_{1:n}$. Note that most of the test statistics under investigation depend on the optimal penalty weights $\hat\llambda(\yy_{1:n})$, which is a function of the observed data. Pilot runs suggest that the variability of the optimal weights is small over bootstrap samples; we therefore treat $\llambda = \hat\llambda(\yy_{1:n})$ as fixed and do not repeat penalty weight selection in the resampling process, which substantially reduces computational time.

Let $S_{ij} = \varsigma_j(Y_{ij})$ be the \textit{MV score} associated with the individual response entry $Y_{ij}$. For continuous log-RT variables and dichotomous items, we simply let $\varsigma_j$ be the identity function and thus $S_{ij} = Y_{ij}$; for unordered polytomous items, however, a customized $\varsigma_j$ function is needed for recoding raw responses to a more meaningful scale (see Section \ref{ss:config} for an example). To assess the lack-of-fit for the simple-structure semiparametric model---in particular the unaccounted dependencies residing in observed MVs, we compute the \textit{residual correlation} statistic
\begin{equation}
  e_{jj'}({\bf y}_{1:n}, \llambda) = r_{jj'} - \rho_{jj'}(\hat\ttheta(\yy_{1:n}, \llambda)).
  \label{eq:rescorr}
\end{equation}
for $j, j' = 1,\dots, m$, $j < j'$. In Equation \ref{eq:rescorr}, $r_{jj'}$ and $\rho_{jj'}$ are the respective sample and model-implied correlations between the $j$th and $j'$th MV scores: The model-implied correlation can be further expressed as
\begin{equation}
  \rho_{jj'} = \frac{\mu_{jj'} - \mu_j\mu_{j'}}
  {\sqrt{(\mu_{jj} - \mu_j^2)(\mu_{j'j'} - \mu_{j'}^2)}},
  \label{eq:expcorr}
\end{equation}
in which we drop the dependency on $\ttheta$ for conciseness. In Equation \ref{eq:expcorr}, the first moment $\mu_j$ can be computed as
\begin{equation}
\mu_j = \int_{\cY_j}\varsigma_j(y)\left[\int_0^1f_j(y|x) dx\right]dy.
  \label{eq:exp1}
\end{equation}
There are three cases when computing the second moment $\mu_{jj'}$: (1) for a single MV, i.e., $j = j'$,
\begin{equation}
  \mu_{jj} = \int_{\cY_j}\varsigma_j(y)^2\left[\int_0^1f_j(y|x) dx\right]dy;
  \label{eq:exp21}
\end{equation}
(2) when $j\ne j'$ but the two MVs load on the same LV,
\begin{equation}
  \mu_{jj'} = \int_{\cY_j\times\cY_{j'}}\varsigma_j(y)\varsigma_{j'}(z)\left[\int_0^1f_j(y|x)f_{j'}(z|x) dx\right]dydz;
  \label{eq:exp22}
\end{equation}
and (3) when the $j$th and $j'$th MVs load respectively on the first and second LVs,
\begin{equation}
  \mu_{jj'} = 
  \int_{\cY_j\times\cY_{j'}}\varsigma_{j}(y)\varsigma_{j'}(z)\left[\int_{[0, 1]^2}f_j(y|x_1)f_{j'}(z|x_2)c(x_1, x_2) dx_1dx_2\right]dydz.
  \label{eq:exp2}
\end{equation}

\subsection{Latent Variable Density and Scores}
\label{ss:lv}
As we have mentioned in Section \ref{ss:cop}, inferences for LVs are made based on the marginally normal $X_{i1}^*$ and $X_{i2}^*$. In particular, we are interested in the strength of association between the two LVs. To this end, we compute the coefficient of determination for predicting ability ($X^*_{i2}$) by slowness ($X^*_{i1}$):
\begin{align}
  \eta^2 =\ & 1 - \frac{\ex\left[ \var(X_{i2}^*|X_{i1}^*) \right]}{\var(X_{i2}^*)}
  = \frac{\var\left[ \ex(X_{i2}^*|X_{i1}^*) \right]}{\var(X_{i2}^*)}\cr 
  =\ & \int_\real \left[\int_\real x_2^*h(x_2^*|x_1^*) dx_2^*  \right]^2h(x_1^*)dx_1^*,
  \label{eq:eta2}
\end{align}
in which $h(x_d^*) = \int_{\real} h(x_1^*, x_2^*)dx^*_{3-d}$, $d = 1, 2$, is the marginal density of $X_{id}^*$ (assumed to be standard normal), and $h(x_2^*|x_1^*) = h(x_1^*, x_2^*) / h(x_1^*)$ is the conditional density of $x_2^*$ given $x_1^*$. Equation \ref{eq:eta2} reduces to the usual coefficient of determination for linear models when $(X_{i1}^*, X_{i2}^*)\t$ follows a bivariate normal distribution. When analyzing real data, we evaluate $\eta^2$ using the estimated LV density, denoted $\hat\eta^2(\yy_{1:n}, \llambda)$; the sampling variability of $\hat\eta^2(\yy_{1:n}, \llambda)$ is again characterized by bootstrapping (Section \ref{ss:fit}).

For each respondent $i$, LV scores can be predicted based on the posterior distribution of $(X_{i1}^*, X_{i2}^*)\t$ given $\yy_i = \yy$ with density
\begin{equation}
  f^h(x_1^*, x_2^*|\yy) = \frac{f(\yy|\Phi(x_1^*), \Phi(x_2^*))h(x_1^*, x_2^*)}{\iint_{\real^2} f(\yy|\Phi(x_1^*), \Phi(x_2^*))h(x_1^*, x_2^*)dx_1^*dx_2^*}.
  \label{eq:post}
\end{equation}
The means of the posterior distribution are often referred to as the expected \textit{a posteriori} (EAP) scores, and the corresponding standard deviations (SDs) gauge the precision of the EAP scores \citep{ThissenWainer2001}. In practice, density functions involved in Equation \ref{eq:post} must be estimated from sample data, which introduces additional uncertainty to scores computed from the estimated posterior. Better precision measures can be obtained from a predictive distribution of LV scores \citep{LiuYang2018a, LiuYang2018b, YangEtAl2012}. Let $\hat f^h(x_1^*, x_2^*|\yy; \yy_{1:n}, \llambda)$ be the estimated posterior density. The bootstrap expectation $\ex\hat f^h(x_1^*, x_2^*|\yy; \bar\YY_{1:n}, \llambda)$ with respect to the (random) bootstrap sample $\bar\YY_{1:n}$ defines a suitable predictive density; the inverse variance of the predictive distribution, which is henceforth referred to as the \textit{predictive precision}, can be conveniently estimated from a collection of bootstrap samples. To set the baseline for assessing the gain in predictive precision, we also consider the marginal posterior density of the ability factor $X_{i2}^*$:
\begin{equation}
  f^h(x_2^*|y_{m_1 + 1}, \dots, y_m) = \frac{\prod_{j=m_1 + 1}^mf(y_j|\Phi(x_2^*))h(x_2^*)}{\int_\real\prod_{j=m_1 + 1}^mf(y_j|\Phi(x_2^*))h(x_2^*)dx_2^*}.
  \label{eq:postmarg}
\end{equation}
Estimated marginal EAP scores and the associated bootstrap predictive precisions can be obtained in a fashion similar to the two-dimensional case.

\section{Data and Analysis Plan}
\label{s:dap}
\subsection{PISA 2015 Mathematics Data}
\label{ss:data}

\begin{table}[!t]
  \centering
  \caption{Descriptive statistics for transformed response time (RT). We applied the based-10 logarithm to the raw RT data and then rescaled the log-transformed variables to $[0, 1]$. For CM155, CM411, CM496, and CM564, summary statistics are computed for the log-transformed RT of the testlets. SD: Standard deviation. Skew: Skewness. Kurt: Kurtosis. CorrTotal: Correlation with the total sum of log-RT.}
  \label{tab:summRT}
  \small
  \begin{tabular}{crrrrrrr}
    \toprule
&CM033Q01 &CM474Q01 &CM155 &CM411 &CM803Q01 &CM442Q02\\
    \midrule
Mean&0.47 &0.42 &0.71 &0.68 &0.53 &0.64\\
SD&0.20 &0.19 &0.12 &0.16 &0.18 &0.16\\
Skew&0.14 &0.54 &$-$0.94 &$-$1.09 &$-$0.02 &$-$0.69\\
Kurt&2.60 &3.08 &5.64 &5.14 &2.89 &4.02\\
CorrTotal&0.41 &0.42 &0.50 &0.50 &0.52 &0.58\\
    \bottomrule
  \end{tabular}\\
  \bigskip
  \begin{tabular}{crrrrrrr}
    \toprule
 &CM034Q01 &CM305Q01 &CM496 &CM423Q01 &CM603Q01 &CM571Q01 &CM564\\
    \midrule
Mean&0.57 &0.59 &0.66 &0.50 &0.69 &0.62 &0.59\\
SD&0.18 &0.16 &0.17 &0.17 &0.18 &0.20 &0.16\\
Skew&$-$0.28 &$-$0.28 &$-$0.98 &0.11 &$-$1.30 &$-$0.92 &$-$0.65\\
Kurt&3.04 &3.46 &4.53 &3.00 &4.84 &3.49 &4.07\\
CorrTotal&0.56 &0.55 &0.45 &0.48 &0.57 &0.52 &0.47\\
    \bottomrule
  \end{tabular}
\end{table}

\begin{table}[!t]
  \centering
  \caption{Descriptive statistics for item responses. For testlets CM155, CM411, CM496, and CM564, the original item response patterns (0, 0), (1, 0), (0, 1), and (1, 1) are recoded to 0, 1, 2, and 3, respectively. P1-3: Observed proportions of response categories 1-3. CorrTotal: Correlation with total score; we apply the scoring function described in Section \ref{ss:config} to the testlet MVs before computing the total score.}
  \label{tab:summR}
  \small
  \begin{tabular}{crrrrrrr}
    \toprule
&CM033Q01 &CM474Q01 &CM155 &CM411 &CM803Q01 &CM442Q02\\
    \midrule
P1 &0.77 &0.66 &0.28 &0.21 &0.26 &0.32\\
P2 & --- & --- &0.11 &0.19 & --- & ---\\
P3 & --- & --- &0.43 &0.29 & --- & ---\\
CorrTotal &0.44 &0.48 &0.49 &0.55 &0.56 &0.58\\
    \bottomrule
  \end{tabular}\\
  \bigskip
  \begin{tabular}{crrrrrrr}
    \toprule
 &CM034Q01 &CM305Q01 &CM496 &CM423Q01 &CM603Q01 &CM571Q01 &CM564\\
    \midrule
P1&0.38 &0.43 &0.07 &0.79 &0.37 &0.41 &0.22\\
P2& --- & --- &0.24 & --- & --- & --- &0.19\\
P3& --- & --- &0.43 & --- & --- & --- &0.27\\
CorrTotal&0.56 &0.31 &0.56 &0.33 &0.45 &0.54 &0.43\\
    \bottomrule
  \end{tabular}
\end{table}
\yl{The data we analyze next came from the PISA 2015 computer-based mathematics assessment \citep{OECD2016}. The test is composed of 17 dichotomously scored items from two mathematics testing clusters (M1 and M2). Similar to the \citet{ZhanEtAl2018}, we only retained cases with complete response entries, leading to a total number of $n = 8606$ observations from 58 countries/economies.}

Among the 17 items, there are four testlets (with item labels starting with CM155, CM411, CM496, and CM564), each of which involves a pair of items. We collapsed the two items within each testlet into a single four-category nominal item: The four categories 0, 1, 2, and 3 indicated the original item response patterns $(0, 0)$, $(1, 0)$, $(0, 1)$, and $(1, 1)$, respectively. The corresponding RT entries were also summed to a single testlet-level RT variable. Accordingly, the number of items involved in the initial fitting is $m_1 = m_2 = 13$, and the number of MVs is $m = 26$. During data preprocessing, we identified a number of extremely small and large RT entries, which are potential outliers and may cause instability in model fitting. Therefore, we excluded for each MV the top and bottom 1\% RT and the associated item response data\footnote{\yl{\citet{ZhanEtAl2018} did not delete any extreme RT entries in their analysis. They performed Bayesian estimation with a somewhat informative prior configuration, which is presumably more stable in the presence of outlying observations.}}. Then we took the base-10 logarithm of the RT variables and rescaled them to the unit interval. Selected descriptive statistics of the final data can be found in \yl{Tables \ref{tab:summRT} and \ref{tab:summR}}.  

\subsection{Analysis Plan}
\label{ss:plan}

As we have mentioned in Section \ref{s:intro}, the data set was analyzed in the previous work by \citet{ZhanEtAl2018} using a parametric simple-structure model. Though we acknowledge the parsimony and thus retain a simple factor structure, our analysis differs substantially from the previous work, because we model MV-LV and LV-LV dependencies in a nonparametric fashion and are able to provide an ultimate assessment for the validity of a simple factor structure in this data set. Once we confirm that the dependencies in the MVs are sufficiently accounted for, we present graphics and statistics based on the fitted model to demonstrate how the respective distributions of item responses and RT are governed by the ability and slowness factors, as well as how ability and slowness covary in the population of respondents.

Major steps of our analysis are outlined as follows.  
\begin{enumerate}[label={\it Step \arabic*.}]
  \item Determine the optimal penalty weights $\hat\llambda(\yy_{1:n})$ by the three-stage procedure described in Section \ref{ss:est}.
  \item Draw $B = 100$ bootstrap samples (i.e., resample with replacement) from the observed data $\yy_{1:n}$ and repeat model fitting in each bootstrap sample with $\llambda = \hat\llambda(\yy_{1:n})$.
  \item Examine the residual correlation statistics (Equation \ref{eq:rescorr}) for all pairs of MVs. Flag a pair if the 90\% two-sided bootstrap CI for the residual correlation fall entirely above 0.1 or below $-0.1$.
  \item Remove problematic items from the test and repeat steps 1-3 until no large residual correlation remains.
  \item Plot the conditional densities of the MVs given the marginally normal LVs (Equation \ref{eq:cdns} with $x = \Phi(x^*)$) and the joint density of the two LVs (Equation \ref{eq:jdnstrans}). Compute estimated $\eta^2$ statistics (Equation \ref{eq:eta2}), EAP scores, and the associated predictive SDs for the scores.
\end{enumerate}

\yl{
  Per the request from two referees, we also report in the supplementary document the empirical risk statistics and density estimates for two parametric models. The first model is a standard baseline model for the joint analysis of item response and RT data, which features linear-normal factor models for log-RT variables, 2PL models for item responses, nominal response models for testlets, and a bivariate normal LV density. Due to the strong parametric assumptions made therein, we do not expect the baseline model to fit the data well. Inspired by the semiparametric fitting, we also specified an updated parametric model with nonlinear factor models with quintic mean functions for log-RT variables, 4PL models for item responses, nominal models for testlets, and a two-component normal mixture density for the LVs. Even though the updated model has yet to attain a fit comparable to the semiparametric model, it reproduces key functional patterns in the semiparametric estimates of the bivariate LV density and the conditional densities for the MVs. Despite being tangential to the specific aims of the present work, these additional analyses exemplify another standard usage of semiparametric/nonparametric models: to provide diagnostic information about model-data fit and to guide model modification.
}

\subsection{Detailed Configuration}
\label{ss:config}

For replicability, we provide all the tuning details involved in our analysis. \yl{PML estimation of the semiparametric simple structure model was implemented in the R package {\tt spfa}, which can be downloaded at \href{https://github.com/wwang1370/spfa}{https://github.com/wwang1370/spfa} and \href{https://cran.r-project.org/web/packages/spfa/index.html}{https://cran.r-project.org/web/packages/spfa/index.html}.}

\paragraph{Estimation}
$K = 13$ B-splines basis functions were used for approximating smooth functions defined on the unit interval. Each log-RT variable was linearly transformed to $[0, 1]$ using the sample minimum and maximum. The reference level for LVs and continuous MVs was set to $x_0 = y_0 = 0.5$; for discrete MVs, the reference level was set to the first response category $y_0 = 0$. We impose likelihood-ratio monotonicity on item CM442Q02 since both its responses and RT show the highest correlations with totals (see Tables \ref{tab:summRT} and \ref{tab:summR}). Intractable integrals appeared in the conditional densities (Equation \ref{eq:cdns}) were approximated by a 21-point Gauss-Legendre quadrature rescaled to the unit interval. The marginal likelihood function (Equation \ref{eq:mlik}) involves a two-dimensional integral over the unit square and was approximated by a tensor-product Gauss-Legendre quadrature. In each fitting, we executed the EM algorithm until the change in the penalized log-likelihood (i.e., Equation \ref{eq:ploglikn}) was less than $10^{-3}$ between consecutive iterations.

\paragraph{Penalty Weight Selection}
We selected the three penalty weights $\lambda_{(c)}$, $\lambda_{(d)}$, and $\lambda_{(g)}$ from the following sequences of decreasing values:
\begin{align*}
  \lambda_{(c)} \in\ & \{10^{-1}, 10^{-2}, \dots, 10^{-6}\},\\
  \lambda_{(d)} \in\ & \{10^1, 10^{-1}, \dots, 10^{-4}\},\\
  \lambda_{(g)} \in\ & \{10^{-2}, 10^{-4}, \dots, 10^{-8}\}.
\end{align*}
The empirical risk was computed by five-fold cross-validation (Equation \ref{eq:risk} with $S = 5$). The smoothest solution within one SE (estimated by Equation \ref{eq:serisk}) from the minimal-risk solution was deemed optimal.

\paragraph{Inference}
Conditional on the optimal penalty weights, we resampled $B = 100$ times with replacement, refit the model in each bootstrap sample, and examine the (approximate) bootstrap distributions of fitted densities and model fit statistics. When computing fit diagnostics and summary statistics, we approximated intractable integrals by the same quadrature systems that were used in parameter estimation. The MV scoring function\footnote{Note that this scoring function was also applied before computing the item-total correlation statistics in Table \ref{tab:summR}.} for testlet responses was defined by $\varsigma_j(0) = 0$, $\varsigma_j(1) = \varsigma_j(2) = 1$, and $\varsigma_j(3) = 2$.

\section{Results}
\label{s:res}

\subsection{Model Fit and Modification}
\label{ss:fitmod}

\begin{figure}[!t]
  \centering
  \includegraphics[width=\textwidth]{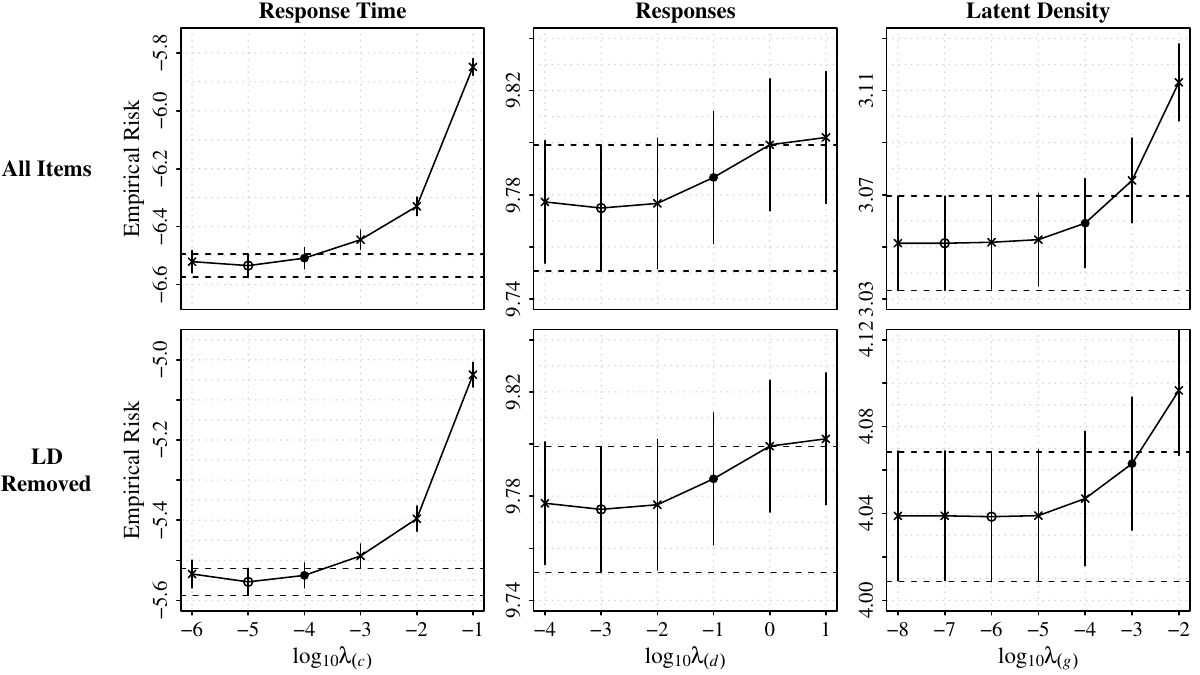}
  \caption{Empirical risks (Equation \ref{eq:risk}) and standard errors (SE; Equation \ref{eq:serisk}). The rows of the graphical table correspond to the initial fitting (with all items) and the updated fitting (without the response time of CM034Q01 and CM571Q01). The columns represent the three stages of penalty weight selection (see Section \ref{ss:est}). Within each panel, empirical risk values are plotted as functions of based-10 log-transformed penalty weights. Vertical bars indicate one SE above and below the empirical risk. The minimized empirical risks are shown as circles, while the optimal solutions determined by the ``one SE rule'' were highlighted as filled dots. \yl{The band formed by two horizontal dashed lines indicates the one-SE region associated with the minimum empirical risk.} Note that the two graphs in the second column are identical: This is because all item responses are retained, and thus we do not need to re-select $\lambda_{(d)}$. LD: Local dependence. $\lambda_{(c)}$, $\lambda_{(d)}$, $\lambda_{(g)}$: Penalty weights for continuous manifest variables (MVs), discrete MVs, and the latent density.}
  \label{fig:risk}
\end{figure}

\begin{figure}[!t]
  \centering
  \includegraphics[width=\textwidth]{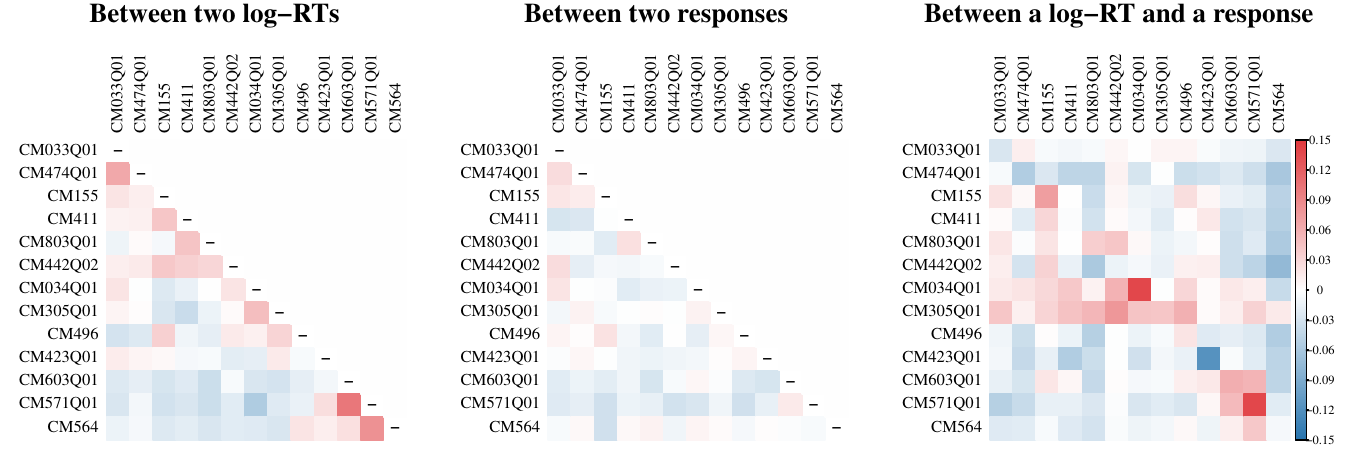}
  \caption{\yl{Residual correlation statistics for the initial fitting of the semiparametric simple-structure model. Left: Residual correlations between two log-transformed response time (log-RT) variables. Middle: Residual correlations between two item/testlet responses. Right: Residuals between a log-RT variable and a item/test response, in which rows represent log-RT variables and columns represent responses. Positive residual correlations are shown in red, while negative residual correlations are shown in blue. A darker color indicates a larger magnitude.}}
  \label{fig:rescorr}
\end{figure}

In the initial fitting of the semiparametric simple-structure model (using all 26 MVs), our cross-validation procedure selects  $10^{-4}$, $10^{-1}$, and $10^{-4}$ as the respective optimal values for $\lambda_{(c)}$, $\lambda_{(d)}$, and $\lambda_{(g)}$. A graphical display of the results can be found in the first row of Figure \ref{fig:risk}. 

Based on a full-data fitting with the optimal penalty weights, we summarize the residual correlation statistics (Equation \ref{eq:rescorr}) for all pairs of MVs in a graphical table (Figure \ref{fig:rescorr}). It is observed that dependencies within RT variables are well explained by the slowness factor, and similarly dependencies within item responses are well explained by the ability factor. The largest residual correlation in the left panel of Figure \ref{fig:rescorr} is 0.1 (between the log-RT of CM571Q01 and CM603Q01) with a 90\% bootstrap CI $[0.08, 0.12]$. In contrast, we identify some non-ignorable residual dependencies between the log-RT and response of the same item (i.e., diagonal entries in the right panel of Figure \ref{fig:rescorr}). The within-item residual correlations reach 0.14 (with a bootstrap CI $[0.12, 0.15]$) for both items CM034Q01 and CM571Q01. We also find a large negative residual correlation for item CM423Q1: The point estimate is $-0.12$, but the associated bootstrap CI $[-0.13, -0.1]$ covers $-0.1$. Meanwhile, the RT-response dependencies are well explained between items: The off-diagonal statistics in the right panel of Figure \ref{fig:rescorr} ranges between $-0.07$ and $0.08$.

Given the above findings, we conclude that a simple factor structure largely suffices for modeling the item responses and RT in the 2015 PISA mathematics data. For two out of 13 items (CM034Q01 and CM571Q01), however, the associations between item-level response speed and accuracy are not fully addressed by individual differences in general processing speed and ability. To be clear of adverse impact caused by unaccounted residual dependencies, we dropped the log-RT variables for items CM034Q1 and CM571Q01 while letting their responses stay, which results in a modified simple-structure model with $m_1 = 11$ continuous MVs and $m_2 = 13$ discrete ones. Steps 1--3 (see Section \ref{ss:plan}) were repeated. The optimal $\lambda_{(c)}$ remains to be $10^{-4}$, whereas the optimal $\lambda_{(g)}$ increases to $10^{-3}$ (see the second row of Figure \ref{fig:risk}); the optimal $\lambda_{(d)} = 10^{-1}$ is retained as no change has been made to the item response variables. There is no more large residual this time. The ranges of the residual correlations are $[-0.04, 0.08]$ among log-RT variables, $[-0.03, 0.02]$ among response variables, and $[-0.11, 0.08]$ across responses and log-RT. Similar to the initial fitting, the only residual correlation beyond $\pm 0.1$ is observed between the response and log-RT of item CM423Q1; however, the 90\% bootstrap CI of the statistic is $[-0.13, -0.09]$ which contains $-0.1$. Therefore, we proceed to interpret the fitted densities based on the updated fitting.

\subsection{Conditional Densities of Manifest Variables}
\label{ss:rescdns}

\begin{figure}[!t]
  \centering
  \includegraphics[width=\textwidth]{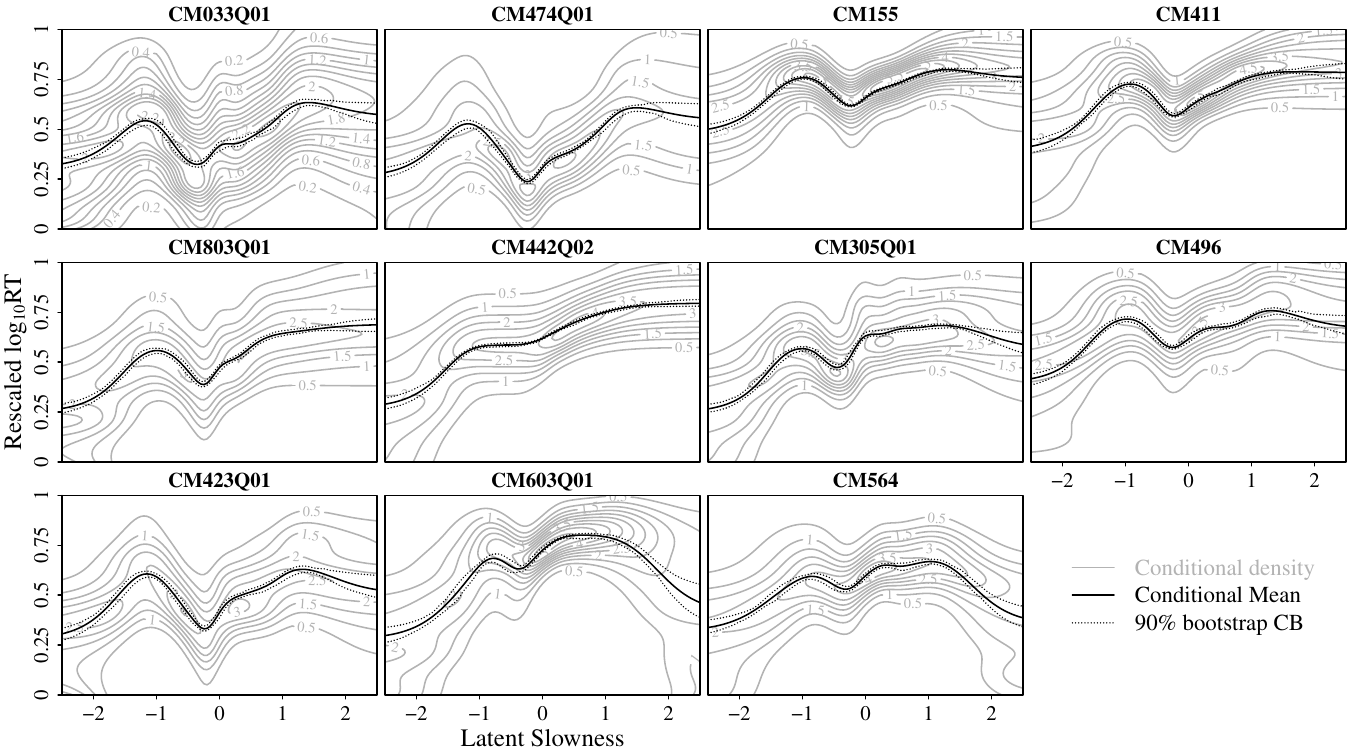}
  \caption{Estimated conditional densities and means for log-10 response time (RT) variables (rescaled to $[0, 1]$). Each panel corresponds to a single item/testlet. Conditional densities of manifest variables given the slowness factor are visualized as contours in gray. Estimated conditional means are superimposed as solid curves in black. Dotted lines represent 90\% bootstrap confidence bands (CBs) for estimated conditional mean curves.}
  \label{fig:cdnsRT}
\end{figure}
\yl{Estimated conditional densities and means of the log-RT variables given the slowness factor are plotted in Figure \ref{fig:cdnsRT}. Two major patterns are of interest here. First, although the high and low ends of the LV scale roughly map onto the longest and shortest RT for a majority of items/testlets, which justifies our decision to label the LV as ``slowness'', the conditional mean function appears to decrease at the high end for all items/testlets except for CM442Q02, on which we impose the monotonicity constraints (Equation \ref{eq:mono}). However, we often cannot distinguish the observed downward trend from a flat one due to large sampling variability, which is manifested by wider bootstrap confidence bands in those areas. For item CM603Q01 and testlet CM564, the downturn at the high end cannot be explained away by sampling variability. It implies that, among slow responders for the first nine items/testlets, the slower they respond to the first nine the faster they tend to response to the last two. The second observation concerns the dips in conditional mean functions when the latent slowness is between $-1$ and 0. Taking sampling variability into account, the dips are not substantial for CM603Q01 and CM564; also recall that the conditional mean function was forced to be non-decreasing for item CM442Q02. As such, the observed dips reflect a negative association between the above triplet and the remaining items/testlets for the subset of respondents whose latent slowness values fall slightly below average.}

\begin{figure}[!t]
  \centering
  \includegraphics[width=\textwidth]{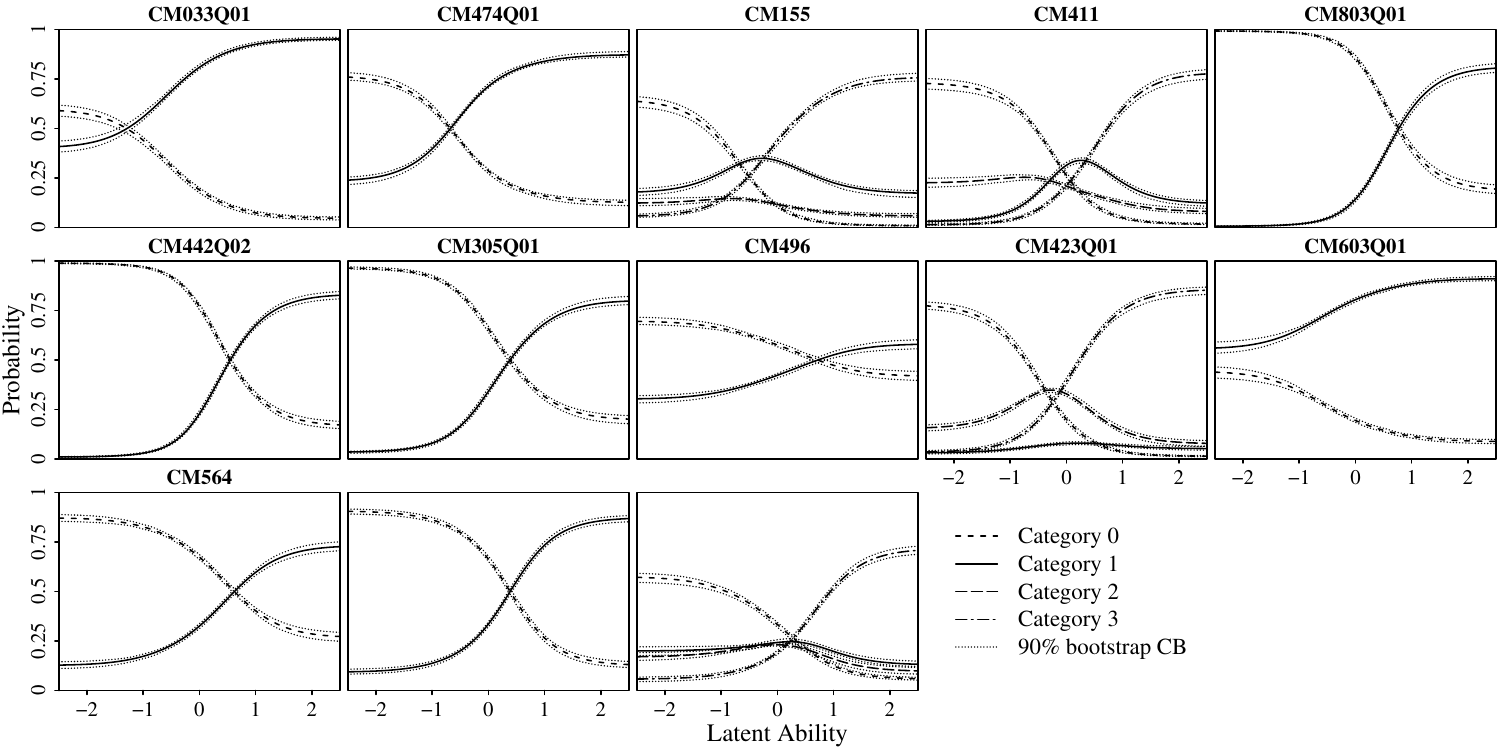}
  \caption{Estimated conditional densities for discrete response variables, also known as item response functions (IRFs). Each panel corresponds to a single item/testlet. Curves for different categories are shown in different line types. Dotted lines represent 90\% bootstrap confidence bands (CBs) for estimated IRFs.}
  \label{fig:cdnsR}
\end{figure}

\yl{Per a referee's request, we also examine the relationship between item-level RT and the ability factor. In our simple structure model, the log-RT variables $Y_{ij}$, $j = 1,\dots, m_1$, do not directly load on the ability factor $X_{i2}^*$. Nevertheless, it remains possible to characterize the predictive distribution $Y_{ij}|X_{i2}^*$ by combining the conditional distribution of the slowness factor given the ability factor, i.e., $X_{i1}^*|X_{i2}^*$, with the conditional distribution $Y_{ij}|X_{i1}^*$ (shown in Figure \ref{fig:cdnsRT}). Such RT-ability associations turn out to be weak in the present data set; detailed results can be found in the supplementary document.}

Estimated item/testlet response functions are displayed in Figure \ref{fig:cdnsR}. Due to the large penalty weight (i.e., $10^{-1}$), the fitted curves are smooth. For dichotomous items, the estimated curves for category 1 (i.e., correct answer) are largely in S-shape and typically have a restricted range (narrow than the entire interval $[0, 1]$). Similarly, estimated testlet response functions for the first and last categories also appear to have (often different) upper asymptotes. Some items, e.g., CM305Q01 and CM423Q01, are poorly discriminating, manifested by relatively flat IRFs.

\begin{figure}[!t]
  \centering
  \includegraphics[width=\textwidth]{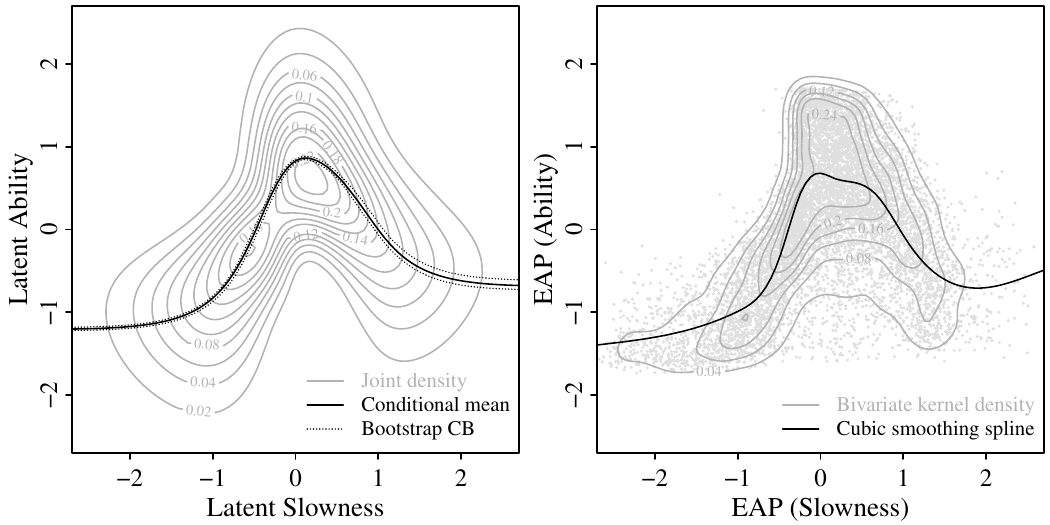}
  \caption{Left: Estimated joint density for the slowness and ability factors (contours in gray) and the conditional mean of ability given slowness (black solid curve). Dotted lines represent 90\% bootstrap confidence bands (CBs) for the estimated conditional mean curve. Right: Scatter plot for the expected \textit{a posteriori} (EAP) scores of ability and slowness. A bivariate kernel density estimate (gray solid contours) and a smoothing spline regression line (black solid curve) are superimposed.}
  \label{fig:dnssco}
\end{figure}

\subsection{Latent Density and Scores}
\label{ss:reslv}
A contour plot for the estimated two-dimensional LV density, which is computed from the estimated B-spline copula density with standard normal marginals (Equation \ref{eq:jdnstrans}), is provided in the left panel of Figure \ref{fig:dnssco}. It is observed that high ability respondents tend to response in a moderate speed, whereas low ability respondents can respond either very rapidly or very slowly. The shape of the density contours is nowhere near elliptical, which calls the standard practice of fitting a bivariate normal LV density into question. A better parameterization of the latent density for this data would be a mixture of two bivariate normals---one with a positive correlation for fast responders (i.e., slowness $<0$) and the other with negative correlation for slow responders (i.e., slowness $>0$). A similar pattern is observed when we plot the ability EAP scores against the slowness EAP scores (right panel of Figure \ref{fig:dnssco}), with an exception that EAP scores tend to be less variable than the true LVs.

To better visualize the relationship between the two latent factors in the population, we also plot the conditional mean of ability given slowness (i.e., the black solid curve in the left panel of Figure \ref{fig:dnssco})---in other words, a nonlinear regression that predicts ability by slowness. The $\eta^2$ statistic (Equation \ref{eq:eta2}) of the population nonlinear regression is 0.45 with a 90\% bootstrap CI $[0.44, 0.47]$, indicating a strong association \citep[][Chapter 9]{Cohen1988}. Stated differently, knowing respondents' processing speed on average reduces the uncertainty (measured by variance) in their mathematics ability by 45\%. Recall that \citet{ZhanEtAl2018} reported a correlation of $-0.2$ between the speed (i.e., the reversal of slowness) and ability factors assuming bivariate normality, which implies $\eta^2=(-0.2)^2 = 0.04$. The divergent conclusion reached by \citet{ZhanEtAl2018} is likely attributed to the restrictive parameterization of their measurement model: They forced the LV density to be bivariate normal and thus failed to capture the nonlinear relationship. In addition, a smoothing spline regression fitted to the EAP scores (i.e., the black sold curve in the right panel of Figure \ref{fig:dnssco})) suggests a similar predictive relationship: The observed multiple $R^2$ statistic is 0.54, even higher than the population $\eta^2$.

\begin{figure}[!t]
  \centering
  \includegraphics[width=\textwidth]{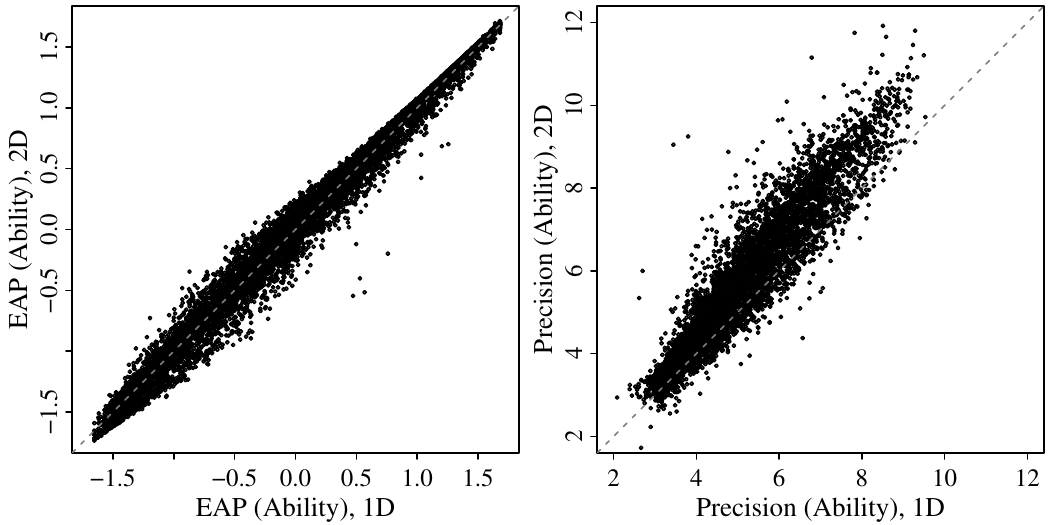}
  \caption{Comparing expected \textit{a posteriori} (EAP) scores for ability (left) and the associated predictive precisions (right) between the one-dimensional (1D) response-only model and the two-dimensional (2D) simple-structural model. In both panels, the dashed diagonal line in gray indicates equality.}
  \label{fig:scoprec}
\end{figure}

As slowness/speed is a useful predictor of ability, it is anticipated that incorporating item-level RT information may improve the precision of IRT scale scores. Inspired by \citet{BolsinovaTijmstra2018}, we compare ability scores from the two-dimensional simple-structure model to those from the unidimensional semiparametric IRT model fitted to only responses in terms of their predictive precision (Section \ref{ss:lv}). It is first noted that the two sets of EAP scores are almost perfectly correlated (sample Pearson's correlation $>$ 0.99; see the left panel of Figure \ref{fig:scoprec}). We then plot the predictive precisions associated with the two sets of EAP scores in the right panel of Figure \ref{fig:dnssco}. Because the test is short and some items (e.g., items CM305Q01 and CM423Q01) have low discriminative power (manifested by flat item response functions), the predictive precisions are not high in general. Pooling across the entire sample, the mean predictive precision based on the unidimensional model is 4.68 with an interquartile range (IQR) $[3.44, 5.71]$, and the median predictive precision based on the two-dimensional simple-structure model is 5.15 with an IQR $[3.57, 6.45]$. That is to say, using the two-dimensional model improves the predictive precision for ability scores by 10.1\% on average. 

\begin{table}
  \color{blue1}
  \centering
  \caption{\yl{Predictive precisions of ability scores in quintile groups. Groups are determined by the slowness (left columns) and ability (right columns) scores computed from the two-dimensional simple-structure model. Avg Prec: Average predictive precision within each group. 1D: One-dimensional model. 2D: Two-dimensional model.}}
  \label{tab:scoprec}
  \begin{tabular}{ccccccccccc}
    \toprule
    & \multicolumn{5}{c}{Quintile Groups (Slowness)}
    & \multicolumn{5}{c}{Quintile Groups (Ability)}\\
    \cmidrule(r){2-6}
    \cmidrule(r){7-11}
    & 1 & 2 & 3 & 4 & 5 & 1 & 2 & 3 & 4 & 5\\
    \midrule
    Avg Prec (1D)      &4.02& 5.01 &4.70  &4.84 &4.82 &3.28& 4.78 &6.55  &5.22 &3.55\\
    Avg Prec (2D)      &4.27& 5.28 &5.03  &5.47 &5.72 &3.46& 5.43 &7.60  &5.63 &3.64\\
    Improvement (in \%)&5.99& 5.31& 7.18 &12.98 &18.64&5.48& 13.47& 15.89 &7.96 &2.52\\
    \bottomrule
  \end{tabular}
\end{table}

\yl{To assess scoring precision at different slowness and ability levels, we split the sample into quintile groups by the slowness and ability EAP scores (from the two-dimensional model), respectively. A group-by-group summary of scoring precisions is provided in Table \ref{tab:scoprec}. When groups are formed by slowness scores, more increases in precision are typically observed in higher quintile groups; the percentage of improvement can be as high as 18.64\% in the fifth quintile group. In contrast, the largest improvement is attained in the middle quintile group ($15.89\%$) when groups are determined by ability scores; the one-dimensional ability scores in the fifth quintile group are almost as precise as the two-dimensional scores.}

\section{Discussion}
\label{s:disc}
In the present paper, we perform a joint factor analysis for item response and RT data from the 2015 PISA mathematics assessment. In line with many previous studies that handled this type of data, our model features a simple factor structure with two LVs: The ability factor is indicated solely by item responses, the slowness factor is indicated solely by log-transformed RT variables, and the two LVs are permitted to covary in the population of respondents. The unique contribution of our work lies in the use of a semiparametric measurement model: We do not impose any restrictive functional forms of dependencies or distributional assumptions above and beyond the simple factor structure. Our model therefore fits the best to the data insofar as a simple factor structure is deemed proper. We approximate the functional parameters in the semiparametric factor model by cubic splines and estimate the resulting coefficients by PML: The penalty weights involved in the objective function are empirically selected via cross-validation. Inferences about model fit statistics and estimated functional parameters are conducted based on (nonparametric) bootstrap.

\subsection{Implications}
\label{ss:imp}

The semiparametric fitting reveals novel patterns that have yet been noticed in the existing literature, which has profound implications on the use of RT information in large-scale educational assessment.

First, a simple factor structure for ability and slowness fits reasonably well to the 2015 PISA mathematics data. Only two pairs of MVs exhibit excessive dependencies that are not well explained by the simple-structure model: Both pairs comprise the response and RT of the same item. Furthermore, including or excluding the RT variables of the two flagged pairs is inconsequential for model-based inferences. Our finding verifies the prevalent psychometric theory that between-person heterogeneity in item response behaviors are reflections of individual differences in ability and general processing speed. However, the existence of within-item local dependence between responses and RT, albeit not influential for the current analysis of the PISA data, should be reassessed in other applications of simple-structure factor models.

Second, commonly used parametric factor models are too simple to fully capture the MV-LV relations. Our semiparametric model implies that the conditional means of log-transformed RT variables are generally increasing but nonlinear functions of the slowness factor; the conditional variances appear to be non-constant for some items too. The most commonly used log-normal RT model, however, implies a linear conditional mean and a constant conditional variance and thus is evidently misspecified. As \citet{LiuWang2022} also reported in that the log-normal RT model fits substantially worse than the semiparametric model in a different empirical example, cautions are advised in choose a suitable measurement model for item-level RT. Meanwhile, a large penalty weight is selected for the semiparametric IRT model, and consequently the fitted IRFs are smooth. While the shapes of the IRFs closely resemble logistic curves, the presence of lower and upper asymptotes hints at a 4PL model \citep{BartonLord1981}, rather than the more popular 1PL and 2PL models in psychometric operations.

Third, the ability and slowness factors are strongly associated, which is probably the most surprising observation since a weak correlation was reported in \citeauthor{ZhanEtAl2018}'s \citeyearpar{ZhanEtAl2018} analysis of the same data. The disparate finding of ours is ascribed to the use of a nonparametric latent density estimator, whereas the LV density is by default assumed to be (multivariate) normal in the vast majority of factor analysis applications. It then merely echoes a well-known fact that overly restrictive assumptions may lead to poorly fitting models and subsequently biased inferences. Diagnostics for non-normal LVs and measurement models equipped with non-parametric LV densities should be added to the routine toolbox for psychometricians. Future research is encouraged to examine the extent to which nonlinear factor models with non-normal latent densities can be beneficial in other assessment contexts.

Fourth, including item-level RT in the measurement model improves the precision of ability scores, which is an expected consequence as the ability factor can be well predicted by the slowness factor. While RT carries additional information about respondents' ability, induced by the association between ability and general processing speed, it remains unclear whether RT should be officially used for scoring purposes in high-stake educational assessment. On the one hand, the joint factor model estimated in the present paper results in about 10\% increase in predictive precisions for ability scores on average. Adaptive tests based on such a joint factor model may need much fewer test items to reach the desired measurement precision, leading to more cost-effective test administrations. On the other hand, the same measurement model may no longer hold once the respondents are aware that response speed somehow affects their performance scores. In the latter case, a re-calibration of the joint factor model and a re-evaluation on the usefulness of RT information are necessary.

\subsection{Limitations}
\label{ss:lim}

There are also a number limitations to be addressed by future investigation.

\yl{First, the selection of penalty weights by multifold cross-validation is time consuming. A referee suggested that computing a one-sample estimate of cross-validation error (e.g., Akaike information criterion; AIC) or a large-sample approximation to the Bayesian marginal log-likelihood (e.g., Bayesian information criterion; BIC) is computationally advantageous. For nonparametric/semiparametric models using penalized smoothing splines, however, we must substitute a properly defined ``effective degrees of freedom (edf)'' for the number of parameters in the usual formulas of those information criteria. The \textit{ad hoc} definition of edf proposed by \citet{LiuEtAl2016} for semiparametric IRT modeling can potentially be extended to the present context; however, the performance of the resulting information criteria in penalty weight selection remains unclear and should be investigated in future work.}

\yl{Second, the sequential selection of multiple penalty weights does not guarantee that a globally optimal combination is found---it was only implemented as a workaround to alleviate the computational burden. Meanwhile, simultaneous selection on an outer-product grid \citep[cf.][]{LiuEtAl2016} suffers from the ``curse of dimensionality'' and may be computationally inviable when the total number of penalty weights to be selected is large. Future research is encouraged to apply and evaluate optimization-based penalty weight selection, such as the ``performance-oriented iteration'' by \citet{Gu1992}, to semiparametric factor analysis. With the aid of optimization-based selection, it is also possible to explore the feasibility of selecting different penalty weights for different MVs, which further enhances the flexibility of the model. 
}

\yl{Third, some of our decisions regarding locally dependent MVs can be refined. While coding each testlet response pattern as a unique category does not lead to any information loss, treating the summed RT within a testlet as a single MV does. In addition, we remove within-item local dependencies between responses and RT by simply excluding the RT variables. Although our treatments suffice for the purpose of the current analysis, it is natural to seek extensions of the proposed model to handle local dependencies in a more elegant way. In our opinion, the best strategy to approach a pair of locally dependent MVs is to directly model their bivariate conditional distribution given the LVs. For example, we may express the joint density of two log-RT variables, say $Y_{ij} = y$ and $Y_{ij'} = z$, given the latent slowness variable $X_{i1} = x$ using a logistic density transform with a three-way fANOVA decomposition \citep{Gu1995, Gu2013}:
  \begin{equation}
    f(y, z|x)\propto\exp\big(g^y(y) + g^z(z) + g^{xy}(x, y) + g^{xz}(x, z) + g^{yz}(y, z) + g^{xyz}(x, y, z) \big).
    \label{eq:fanova3}
  \end{equation}
  Equation \ref{eq:fanova3} involves six functional components, each of which can be approximated via basis expansion under suitable side conditions. Despite the straightforward formulation, simultaneous estimation of a large number of functional parameters proves to be computationally challenging.
}

\yl{Fourth, a referee made an important point that the residual correlation statistic (Equation \ref{eq:rescorr}) only captures linear dependencies, which does not rule out the existence of nonlinear residual dependencies and is a major limitation of our diagnostic procedure. There exist various measures for nonlinear associations: Recent example include the Hellinger correlation \citep{GeenensLdM2022} and the Wasserstein dependence coefficient \citep[][see also \citeauthor{Chatterjee2022}, \citeyear{Chatterjee2022} for a review]{MordantSegers2022}. However, those measures are often less intuitive to interpret as no common rules of thumb have been developed. As an alternative, one may fit an extended semiparametric factor model with bivariate conditional densities (Equation \ref{eq:fanova3}) and identify nonlinear dependencies from graphical displays of estimated conditional densities.}

\yl{Fifth, the proposed semiparametric factor model can be generalized in a number of ways. Sometimes, multiple latent constructs are simultaneously measured by an instrument (e.g., personality assessment); hence, a joint factor analysis of responses and RT for those measures involves at least three LVs. Such extensions of the current semiparametric simple-structure model suffers from a two-fold ``curse of dimensionality'': The number of tensor-product basis functions grows exponentially when the dimension of a functional parameter's domain increases, and the number of tensor-product quadrature points for likelihood approximation also increases exponentially as the dimension of LVs increases. While the EM algorithm with numerical quadrature can be replaced by stochastic approximation \citep{Cai2010a, Cai2010b, GuKong1998} to handle models with higher-dimensional LVs, reduced fANOVA parameterizations for conditional densities \citep{Gu1995, Gu2013} and hierarchical formulations of B-spline copula \citep{KauermannEtAl2013} are handy for constructing economical approximations of multivariate functional parameters.}

Sixth, resampling based procedures (e.g., bootstrap) are time consuming even if parallel processing via OpenMP \citep{OpenMP} is enabled in the current implementation of PML estimation. For parametric models, inferential procedures based on large-sample approximations fares more computationally efficient. However, it is generally more difficult to prove large-sample results for semiparametric/nonparametric models as the functional parameters are infinite dimensional. Theoretical foundations on the asymptotic theory for semiparametric/nonparametric measurement models have yet been established and are left for future research.

\yl{Last but not least, we emphasize that semiparametric approaches are better suited for analyses that are exploratory and data-driven in nature. There are also scenarios in which confirmatory and theory-driven model building is preferred: For instance, when the test is designed based on cognitive theory and administered in a controlled laboratory setting \citep[e.g., the well known ``mental rotation'' example in the RT literature;][]{BorstEtAl2011}. One prominent example of theory-driven psychometrics is the integration of diffusion decision models with factor analysis \citep[e.g.,][]{KangEtAl2022, KangEtAl2023a, KangEtAl2023b}. Data-driven semiparametric models and theory-driven parametric models are both important yet mutually distinct tools to advance psychometricians' understanding in the role of processing speed in test-taking behavior.}

% supplemental file
\renewcommand\appendixname{Supplementary Appendix}
\appendix
\renewcommand{\thesection}{\Alph{section}}
\renewcommand{\theequation}{S\arabic{equation}}
\renewcommand{\thetable}{S\arabic{table}}
\renewcommand{\thefigure}{S\arabic{figure}}
\setcounter{section}{1}
\setcounter{subsection}{0}
\section{Parametric Fittings}
\subsection{Baseline Model}
The baseline parametric model features simple linear normal factor analysis models for log-transformed response time (RT) variables, standard item response theory (IRT) models for item responses, and a bivariate normal density for the latent slowness and ability. Given the latent slowness $X_{i1}$\footnote{We adopt the same notational convention as we used in the main document.}, the log-RT variable $Y_{ij}\in\real$, $j = 1,\dots, m_1$, is assumed to be normally distributed. The conditional density of $Y_{ij} = y$ given $X_{i1} = x$ is denoted
\begin{equation}
  f_j(y|x) = \phi(y; \mu_j(x), \sigma_j^2),
  \label{eq:cdns1}
\end{equation}
in which $\phi(\cdot; \mu, \sigma)$ is a generic notation for the density of ${\cal N}(\mu, \sigma^2)$, and the mean function
\begin{equation}
  \mu_j(x) = \zeta_j + \gamma_jx
  \label{eq:mean1}
\end{equation}
is assumed to be linear. There are three free parameters for each continuous MV: the intercept $\zeta_j$, the common factor loading $\gamma_j$, and the unique variance $\sigma_j^2$. Given the latent ability $X_{i2}$, the item response function (IRF) for a discrete item response $Y_{ij}\in\{0, \dots, C_j - 1\}$, $j = m_1 + 1, \dots, m$, can be expressed as
\begin{equation}
  \pr\{Y_{ij} = y|X_{i2} = x\} = \left\{\begin{array}{ll}
      \displaystyle\frac{1}{1 + \sum_{c=1}^{C_j - 1}\exp(\zeta_{jc} + \gamma_{jc}x)},&y=0,\\[14pt]
      \displaystyle\frac{\exp(\zeta_{jy} + \gamma_{jy}x)}{1 + \sum_{c=1}^{C_j - 1}\exp(\zeta_{jc} + \gamma_{jc}x)},& y = 1, \dots, C_j - 1,
  \end{array}\right.
  \label{eq:irf1}
\end{equation}
in which $\zeta_{jc}$ and $\gamma_{jc}$, $c = 1,\dots, C_j - 1$, are referred to as the intercept and slope parameters. When $K = 2$, Equation \ref{eq:irf1} reduces to the IRF of the two-parameter logistic (2PL) model, which is specified for dichotomous responses in the baseline model. Finally, the joint density for the two latent variables (LVs) $(X_{i1}, X_{i2})\t$ is assumed to be bivariate normal with density
\begin{equation}
  h(x_1, x_2) = \phi\left(\begin{bmatrix}
      x_1\\x_2
    \end{bmatrix}; \begin{bmatrix}
      0\\0
    \end{bmatrix}, \begin{bmatrix}
      1 & \\
      \sigma_{21} & 1\\
\end{bmatrix}\right),
  \label{eq:latent1}
\end{equation}
in which $\phi(\cdot; \boldsymbol\mu, {\bf\Sigma})$ stands for a bivariate normal density with a mean vector $\boldsymbol\mu$ and a covariance matrix $\bf\Sigma$. Both $X_{i1}$ and $X_{i2}$ are marginally distributed as ${\cal N}(0, 1)$; therefore, the conditional MV densities in the baseline model are directly comparable with those in the semiparametric model.

\subsection{Updated Model}

The semiparametric fitting reported in the main document indicates that a more flexible model is needed to capture the complex dependencies among the observed item responses and RT. Therefore, we proceed to modify the baseline model as follows. For the log-RT variables, we specify a nonlinear normal factor model: While the conditional distribution $Y_{ij}|X_{i1}$ is still characterized by Equation \ref{eq:cdns1}, the mean function is now a quintic polynomial of form 
\begin{equation}
  \mu_j(x) = \zeta_j + \gamma_{j1}x + \gamma_{j2}x^2 + \gamma_{j3}x^3 + \gamma_{j4}x^4 + \gamma_{j5}x^5.
  \label{eq:mean2}
\end{equation}
To match the semiparametric setup, we require the mean log-RT function of item CM442Q02 to be monotonically increasing. This is achieved by imposing positivity on the derivative of Equation \ref{eq:mean2} via the parameterization
\begin{equation}
  \mu_j'(x) = \omega_j\left[1 - 2\upsilon_{j1}x + (\upsilon_{j1}^2 + \tau_{j1})x^2\right]\left[1 - 2\upsilon_{j2}x + (\upsilon_{j2}^2 + \tau_{j2})x^2\right],
  \label{eq:mean3}
\end{equation}
in which $\omega_j,\tau_{j1},\tau_{j2}>0$ \cite{Elphinstone1985,FalkCai2016a}. The corresponding quintic polynomial coefficients can be obtained by straightforward algebra. While testlet responses are still modeled by Equation \ref{eq:irf1} in the updated model, dichotomous item responses are now modeled by the four-parameter logistic (4PL) model with IRF
\begin{equation}
  \pr\{Y_{ij} = 1|X_{i2} = x\} =\varpi_j + \frac{(\varrho_j - \varpi_j)\exp(\zeta_j + \gamma_jx)}{1 + \exp(\zeta_j + \gamma_jx)},
  \label{eq:irf2}
\end{equation}
in which $\varpi_j, \varrho_j\in[0, 1]$, $\varpi_j<\varrho_j$, are the lower- and upper-asymptote parameters. Finally, the latent slowness and ability are assumed to jointly follow a mixture of two independent bivariate normal distributions:
\bgroup\small
  \begin{equation}
  h(x_1, x_2) = \pi\phi\left(\begin{bmatrix}
      x_1\\x_2
    \end{bmatrix}; \begin{bmatrix}
      1\\0.5
    \end{bmatrix}, \begin{bmatrix}
      0.5^2 & \\
      \sigma_{21}^{(1)} & 0.75^2\\
\end{bmatrix}\right) + (1 - \pi)\phi\left(\begin{bmatrix}
      x_1\\x_2
    \end{bmatrix}; \begin{bmatrix}
      \mu_1^{(2)}\\ \mu_2^{(2)}
    \end{bmatrix}, \begin{bmatrix}
      \sigma_{11}^{(2)} & \\
      \sigma_{21}^{(2)} & \sigma_{22}^{(2)}\\
  \end{bmatrix}\right).
  \label{eq:latent2}
\end{equation}\egroup
In Equation \ref{eq:latent2}, the free parameters are the mean and (co)variance parameters $\mu_1^{(2)}$, $\mu_2^{(2)}$, $\sigma_{11}^{(2)}$, $\sigma_{21}^{(2)}$, and $\sigma_{22}^{(2)}$ for the second latent class, the covariance parameter $\sigma_{21}^{(1)}$ for the first latent class, and the class membership probability $\pi$. The means and variances for the first latent class are arbitrarily set in order to identify the LV scale. Note that the joint density characterized by Equation \ref{eq:latent2} does not have ${\cal N}(0, 1)$ marginals. To facilitate comparison with the semiparametric fitting, let 
\begin{equation}
  X_{id}^* = \Phi^{-1}(H_d(X_{id})),\ d = 1, 2,
  \label{eq:trans}
\end{equation}
in which $H_d$ stands for the $d$th marginal distribution function under the joint density (Equation \ref{eq:latent2}). By the probability integral transform, both $X_{i1}^*$ and $X_{i2}^*$ are ${\cal N}(0, 1)$ variates. In the sequel, all the density estimates based on the updated model are plotted for the standard normal $X_{i1}^*$ and $X_{i2}^*$ unless otherwise specified.

\subsection{Results}
Parameters of the two parametric models were estimated using an R implementation of the expectation-maximization (EM) algorithm. The intractable marginal likelihood functions were approximated by an outer-product rectangular quadrature, in which there are 21 equally spaced points ranging from $-5$ to 5 per dimension. The EM algorithm was terminated when the log-likelihood change between consecutive iterations is less than 0.001. 

\begin{figure}[!t]
  \centering
  \includegraphics[width=\textwidth]{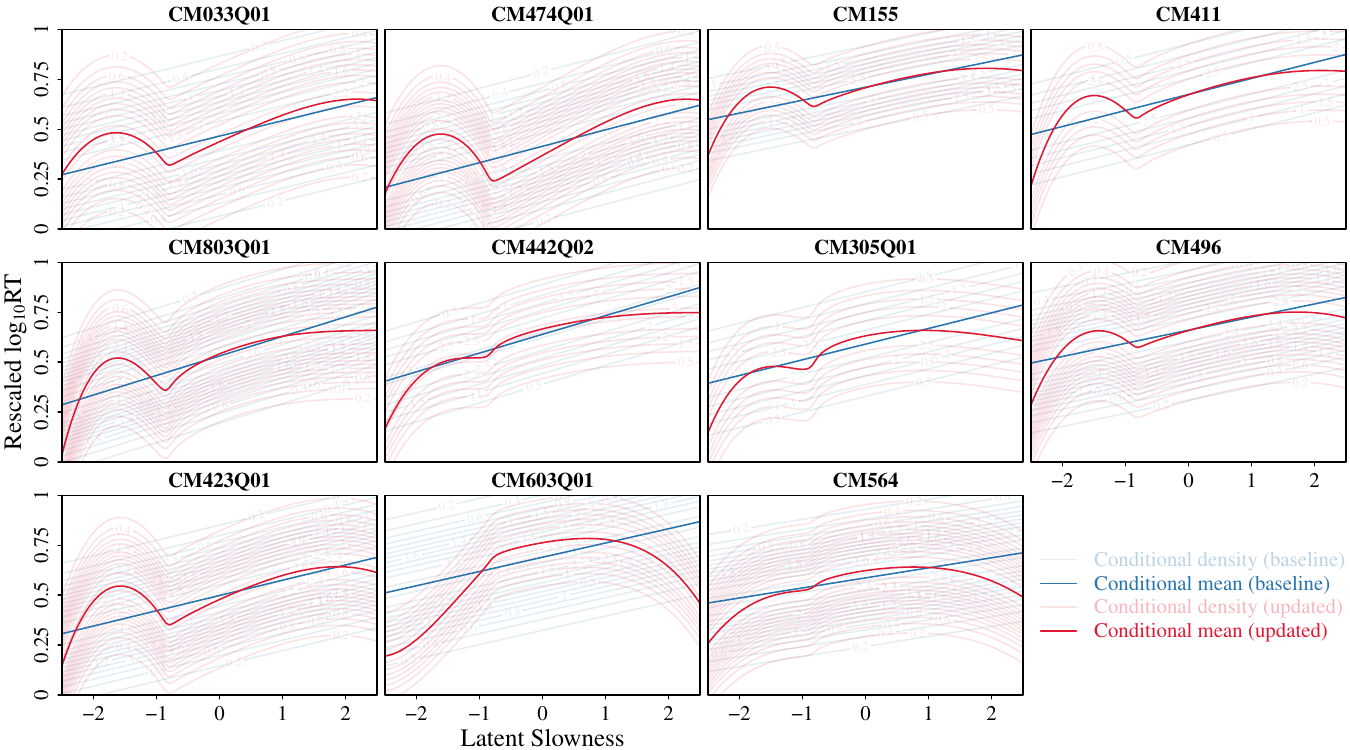}
  \caption{Estimated conditional densities and means for log-10 response time (RT) variables (rescaled to [0, 1]) from the two additional parametric models. Each panel corresponds to a single item/testlet. Conditional densities of manifest variables given the slowness factor are visualized as contours in lighter colors. Estimated conditional means for both the are superimposed as solid curves in darker colors.}
  \label{fig:cdnsRTpar}
\end{figure}

We performed five-fold cross-validation and computed the empirical risk defined in a fashion similar to Equation \ref*{eq:risk}. The empirical risk for the baseline model is 5.17 with a standard error (SE) of 0.16; the empirical risk for the updated model is 4.77 with an SE of 0.15. Compared to the empirical risk of the semiparametric model, i.e., 4.06 with an SE of 0.03 as shown in the last panel of Figure \ref*{fig:risk}, the empirical risks for the two parametric models are substantially higher and thus imply poorer fit.

The estimated conditional densities and conditional mean functions for the log-RT variables are depicted in Figure \ref{fig:cdnsRTpar}. On the one hand, the linear estimates (blue) of the conditional mean functions are quite different from the (transformed) quintic estimates (red) and the semiparametric fitting (Figure \ref*{fig:cdnsRT} in the main document), which contributes to the poor model-data fit of the baseline model. On the other hand, the (transformed) quintic and the semiparametric fittings are more or less aligned in their basic shape. The funky shapes of the quintic fittings around $X_{i1}^* = -1$ are consequences of the LV transformation (Equation \ref{eq:trans}).

\begin{figure}[!t]
  \centering
  \includegraphics[width=\textwidth]{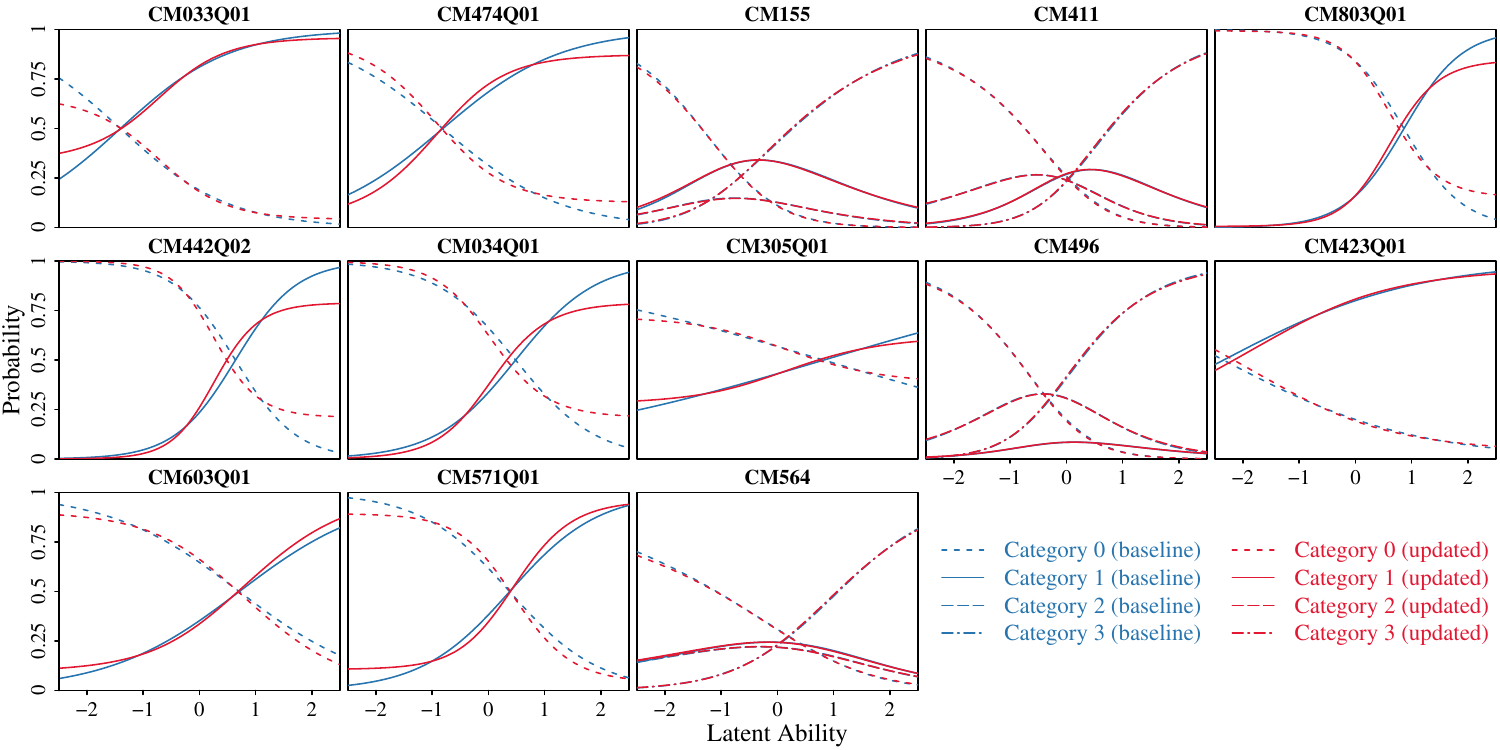}
  \caption{Estimated conditional densities for discrete response variables, also known as item response functions (IRFs), from the two additional parametric models. Each panel corresponds to a single item/testlet. Curves for different categories are shown in different line types. The two models are distinguished by different colors.}
  \label{fig:cdnsRpar}
\end{figure}

The estimated IRFs for item/testlet responses are displayed in Figure \ref{fig:cdnsRpar}. For all the testlets and a majority of dichotomous items, there is little discrepancy between the two parametric fittings, which also closely resemble the semiparametric fitting (Figure \ref*{fig:cdnsR} in the main document). Exceptions include the non-trivial upper asymptotes obtained for items CM803Q01, CM442Q02, and CM031Q01, and the non-trivial lower asymptote for item CM033Q01. 

\begin{figure}[!t]
  \centering
  \includegraphics[width=\textwidth]{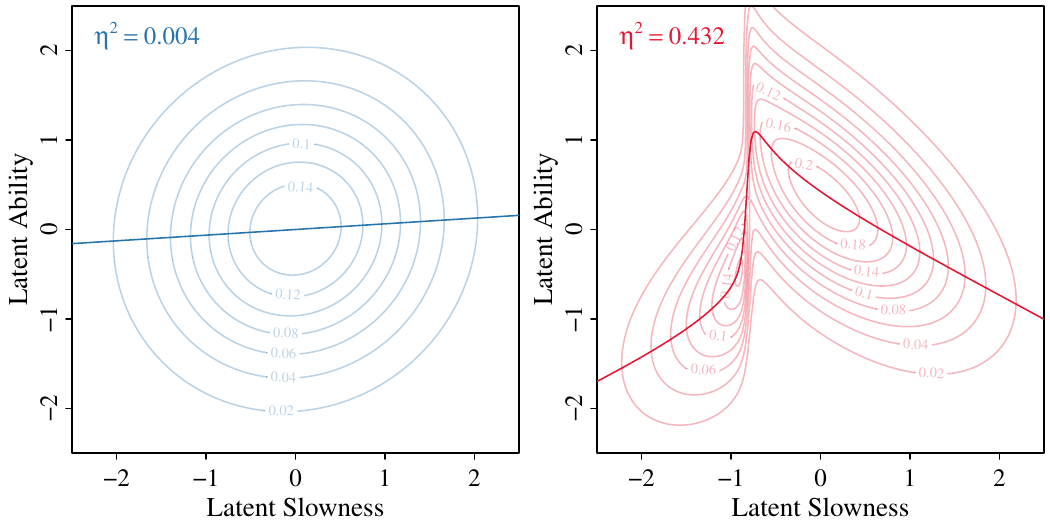}
  \caption{Estimated joint density for the slowness and ability factors (contours in lighter colors) and the conditional mean of ability given slowness (solid curves in darker colors). The population $\eta^2$s for predicting latent ability by latent slowness are printed at the upper-left corner of each panel. Left: Baseline model. Right: Updated model.}
  \label{fig:dnsGpar}
\end{figure}

In Figure \ref{fig:dnsGpar}, the estimated bivariate LV densities from the two parametric models are displayed and contrasted. In the baseline model, the latent slowness and ability are forced to follow a bivariate normal distribution. We therefore obtain a nearly zero inter-factor correlation (about 0.06) and thus a very weak $\eta^2$ (less than 0.01). The two-component mixture density in the updated model allows us to capture the nonlinear relationship between the latent slowness and ability. The resulting $\eta^2$ value for the updated model (0.43) is also close to what we have obtained in the semiparametric fitting (0.45), indicating that the latent ability can be effectively predicted by the latent slowness. Nevertheless, it is noticed that the estimated bivariate LV density in the updated model, after being transformed to have standard normal marginals, still differs from the estimate in the semiparametric fitting (Figure \ref*{fig:dnssco}). This discrepancy is likely resulted from the limited flexibility of the two-component mixture density.

\clearpage

\section{Relationship Between Log-RT and Ability Based on the Semiparametric Model}

\begin{figure}[H]
  \centering
  \includegraphics[width=\textwidth]{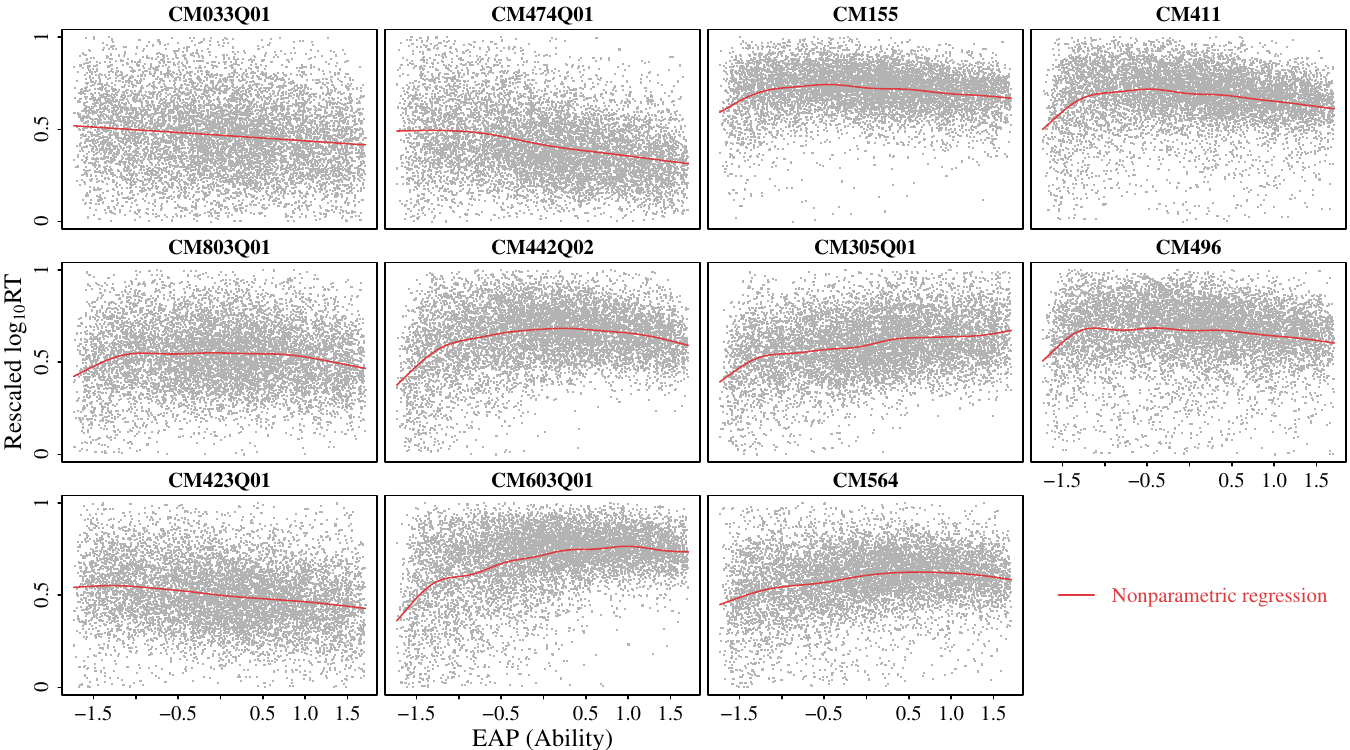}
  \caption{Rescaled log-transformed response time (RT) plotted as functions of expected {\it a posteriori} (EAP) scores for ability. Each panel corresponds to a single item/testlet. Smoothing spline regression lines are superimposed to visualize the trend.}
  \label{fig:RTability1}
\end{figure}

In Figure \ref{fig:RTability1}, the item-level log-RT variables are plotted against the expected {\it a posteriori} (EAP) scores of the ability factor computed from the semiparametric simple-structure model. Inverted U-shape relationships are observed for a number of items, and monotonic relationships are observed for the rest. However, the association between the log-RT and the ability scores is in general weak.

\begin{figure}[H]
  \centering
  \includegraphics[width=\textwidth]{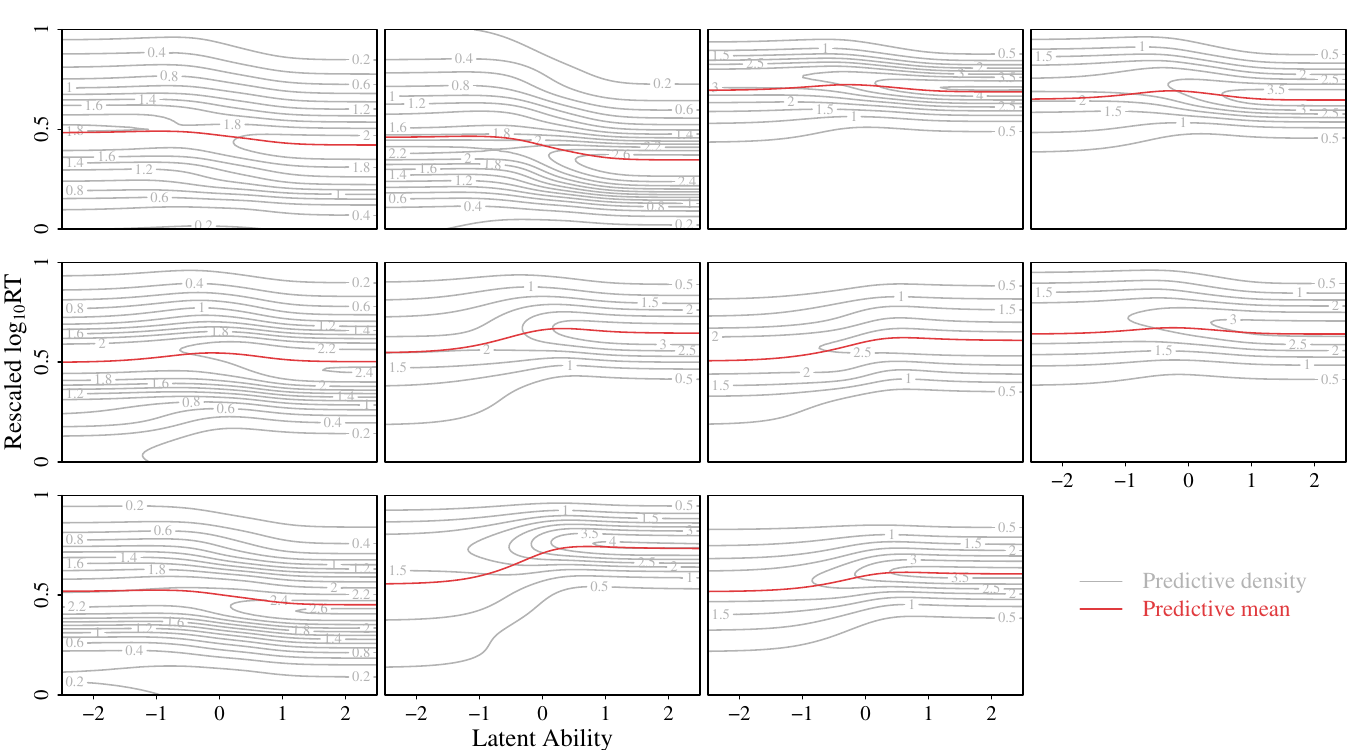}
  \caption{Predictive densities (gray contours) and mean functions (red solid lines) for log-transformed response time (RT) by latent ability. Each panel corresponds to a single item/testlet.}
  \label{fig:RTability2}
\end{figure}

It is also possible to describe the relationship between an item-level log-RT variable and the latent ability by a predictive distribution and the corresponding mean function. In the simple-structure model, the log-RT variables $Y_{ij}$, $j = 1,\dots, m_1$, are not directly related to the latent ability $X_{i2}^*$. But because we can predict the latent slowness $X_{i1}^*$ by $X_{i2}^*$ and $X_{i1}^*$ is indicated by the log-RT variables, we can examine the predictive distribution of $Y_{ij} = y$ given $X_{i2}^* = x_2^*$ characterized by the density
\begin{equation}
  p_j(y|x_2^*) = \int_\real f_j(y|x_1^*)h(x_1^* | x_2^*)dx_1^*,
  \label{eq:preddist}
\end{equation}
in which $h(x_1^*|x_2^*) = h(x_1^*, x_2^*) / h(x_2^*)$ is the conditional density of $X_{i2}^*$ given $X_{i1}^*$. The predictive mean of $Y_{ij}$ at $X_{i2}^* = x_2^*$ is then
\begin{equation}
  \ex(Y_{ij}|X_{i2}^* = x_2^*) = \int yp_j(y|x_2^*)dy.
  \label{eq:predmean}
\end{equation}
We plot Equations \ref{eq:preddist} and \ref{eq:predmean} in Figure \ref{fig:RTability2}. Both Figures \ref{fig:RTability1} and \ref{fig:RTability2} suggest that item-level log-RT variables cannot be reliably predicted by the latent ability.

\clearpage

% bib
\setcounter{secnumdepth}{0}
\bibliographystyle{apacite}
\bibliography{RTPISA2015-1stRev}

\end{document}